\documentclass[10pt,aps,pre,twocolumn,showpacs]{revtex4}
\usepackage{amsmath}
\usepackage{amssymb}
\usepackage{graphicx}
\usepackage{dcolumn}
\usepackage{bm}
\usepackage[below]{placeins}

\begin{document}
\title{Quantum coherence in the dynamical excitation, ionization, and
decaying of neon gas induced by X-ray laser}
\author{Yongqiang Li$^{1}$, Cheng Gao$^{1}$, Wenpu Dong$^{1}$, Jiaolong Zeng$^{1}$ and Jianmin Yuan$^{1,2}$}
\affiliation{$^{1}$Department of Physics, National University of Defense Technology, Changsha 410073, P. R. China \\
$^{2}$IFSA Collaborative Innovation Center, Shanghai Jiao Tong
University, Shanghai 200240, P. R. China}
\email{jmyuan@nudt.edu.cn}


\begin{abstract}
We develop a large scale quantum master equation approach to
describe dynamical processes of practical open quantum systems
driven by both coherent and stochastic interactions by including
more than one thousand true states of the systems, motivated by the
development of highly bright and fully coherent lasers in the X-ray
wavelength regime. The method combines the processes of coherent
dynamics induced by the X-ray laser and incoherent relaxations due
to spontaneous emissions, Auger decays, and electronic collisions.
As examples, theoretical investigation of {\it real} coherent
dynamics of inner-shell electrons of a neon gas, irradiated by a
high-intensity X-ray laser with a full temporal coherence, is
carried out with the approach. In contrast to the rate equation
treatment, we find that coherence can suppress the multiphoton
absorptions of a neon gas in the ultra-intense X-ray pulse, due to
coherence-induced Rabi oscillations and power broadening effects. We
study the influence of coherence on ionization processes of neon,
and directly prove that sequential single-photon processes for both
outer- and inner-shell electrons dominate the ionizations for the
recently typical experiments with a laser intensity of
$\approx10^{18}$ ${\rm W/cm^2}$. We discuss possible experimental
implementations such as signatures for coherent evolution of
inner-shell electrons via resonance fluorescence processes. The
approach can also be applied to many different practical open
quantum systems in atomic, quantum optical, and cold matter systems,
which are treated qualitatively by a few-level master equation model
before.
\end{abstract}

\pacs{Subject Areas: Atomic and Molecular Physics, Optics, Computational Physics}
\date{\today}
\maketitle



\section{Introduction}
Master equation approach is a standard technique for open quantum
systems~\cite{RPL1} and a successful theory in descriptions of
light-matter interactions, such as in condensed matter
physics~\cite{RPL2}, chemistry and biology~\cite{PRL3}, quantum
optics~\cite{PRL4}, and ultracold gases~\cite{J. Cirac}. The master
equation is quite general and encompasses various physical
phenomena, as long as these phenomena share common physical
mechanisms, i.e. the interplay between coherence and dissipation.
Traditionally, these kinds of systems are treated within the
framework of few-level models~\cite{J. Cirac, A. Imamoglu, S. M.
Cavaletto}. With the development of highly bright
lasers~\cite{Seres}, however, the few-level models lose the
possibility for describing complex open systems, since energy here
is deposited in a broad range and relaxes in a vast number of decay
channels. In this case, new challenges appear for real dynamics of
these complex systems, and a large-scale simulation is inevitable.
Here, one open issue, related to coherence effects in dynamical
processes of complex systems, is still unknown.

Coherence plays an important role in describing correlation
properties of quantum matters and understanding quantum phenomena
including lasing~\cite{Maiman}, Fano shape~\cite{Fano},
superconductivity, superfluidity and Bose-Einstein
condensate~\cite{Annett,BEC}, and novel phenomena arising from
quantum optics~\cite{Harris} and attosecond physics~\cite{Plaja}. To
study these coherence-induced quantum features, dramatic success has
also been achieved in preparing and controlling the coherent
dominant systems with suppressed dissipations, such as
electromagnetically induced transparency in quantum
optics~\cite{Harris2} and superfluid-Mott phase transition in
condensed matter physics~\cite{SF-MI}. Recently, exploring coherence
effects of complex systems has been
arguably one of the most important topics in physics, inspired by
generating X-ray laser with extremely high brightness and fully
temporal coherence~\cite{Seres}, such as the Linac Coherent Light
Source (LCLS)~\cite{P. Emma_2010}, where one opened a new era of
exploring the interaction of high-intensity X rays with complex
systems on femtosecond (fs) timescales. Unfortunately, a typical
feature of X-ray-matter interactions is the rapid decay processes,
due to the vast relax channels. As a result, it is inevitable to
investigate the interplay between coherence-induced effects and
dissipations in X-ray-matter systems.

Here, the X-ray free electron laser, with an ultrashort pulse
duration and a high peak brilliance, provides the possibility for
the study of physical, chemical and biological properties which are
never accessed experimentally before~\cite{F. Lehmkuhler}. For
example, a series of pioneer experiments~\cite{Young, M. Hoener, J.
Cryan, G. Doumy, H. Thomas, N. Rohringer1, B. Rudek, B. Nagler, S.
M. Vinko, H. Fukuzawa} focus on understanding of X-ray-matter
interactions, including nonlinear photoionizations, hole
relaxations, Auger electron distributions, and X-ray emission
spectrums in atoms, molecules and solid materials. To simulate these
experiments, a semiclassical description of radiation-field coupled
systems, Einstein's rate equation approach~\cite{N. Rohringer2, O.
Ciricosta, W.-J. Xiang}, has recently been extended to the regime of
X-ray-matter interactions for gaining fundamental insight into the
fast decayed systems, by including photoexcitation and ionization,
electron impact excitation and ionization, Auger decay, and their
reverse processes. In Einstein's rate equation model, all the
absorption and emission processes are treated with transition
probability and the coherence between different levels is
neglected, which are good approximations for either incoherent light
fields or dominant decay processes of the systemic
coherence~\cite{Robert}. The agreement that one finds between
theoretical predictions~\cite{W.-J. Xiang} and
experiments~\cite{Young} at relatively low atomic number densities,
verifies the capability of the model to simulate X-ray-matter
interactions. The physical reason is that the free-electron laser
beam has limited temporal coherence with each pulse consisting of a
random number of incoherent intensity spikes in a fs duration.

One big challenge, however, which has not been overcome yet, is the
temporal coherence of X-ray free-electron laser generated from
self-amplified spontaneous emissions from electron
beams~\cite{SASE}. One goal of the X-ray free-electron laser is to
enhance temporal coherence of the field, and the quest for improved
X-ray sources, based on the free electron laser, is being tackled
with several techniques~\cite{P. Emma, Y. Ding, N. R. Thompson, Y.
Wang, J. Zhao, K. Lan, S. Jacquemot, N. Rohringer1, N. Rohringer3}.
In fact, the new X-ray pulse with improved temporal
coherence~\cite{N. Rohringer3} is more suitable to locally deposit
energy and prepare electronic states, study dynamical properties via
photon correlation spectroscopy, and image biological specimens and
long-rang orders in liquid and condensed matters~\cite{F.
Lehmkuhler}. Recently, one generates the first successful coherent
free-electron laser radiation pulses in the soft X-ray regime, based
on the seeding experiment~\cite{S. Ackermann}, which experimentally
provides the possibility for investigating the interplay between
coherence and dissipation in X-ray-matter systems. Then the crucial
issue, related to an X-ray laser with an improved temporal
coherence, is how to model the ultrafast dynamics of the
X-ray-matter systems and understand the underlying physics.  To
obtain a more precise description of dynamical mechanics, in
general, we need a quantum mechanical tool for simulations of its
time evolution, such as time-dependent Schr\"odinger equation. In
contrast to the case of low-Z species in dilute gases~\cite{Kenneth C. Kulander, M.
Lewenstein, M. Protopapas, Z.-X. Zhao, J. Zhao_2008, B. Zhang},
time-dependent Schr\"odinger equation for complex systems is
difficult to tackle directly, due to electron-electron correlations,
collision processes, and spontaneous and Auger decay processes.
Approximations for the X-ray-matter systems are inevitable, such as
master equation approach. In the master equation approach, one can
include the important microscopic processes as much as possible and
investigate the interplay between coherence and dissipations in the
X-ray-matter systems. Instead of a few-level simulations~\cite{A.
Imamoglu, S. M. Cavaletto, H.-P. Breuer, B. R. Mollow, R. Brewer, P.
Zoller, P. Zoller_1993, J. Cirac}, one need to simulate {\it real}
dynamics of the complex X-ray-matter systems based on large scale
simulations, due to the vast decay channels induced by the intense
X-ray laser. As far as we know, this problem is never explored in
the framework of quantum master equation approach before, coherent
dynamics of the complex systems is still unclear, and it is still an
open issue whether new phenomena arise from coherence effects in the
ultrafast decayed systems.

These questions motivate our study in this paper to establish a
general method for describing coherent dynamics of the X-ray-matter
systems in the framework of master equation approach. In principle,
master equation approach can deal with different kinds of dynamics
of the intense X-ray-matter systems by including the microscopic
processes due to photons, electrons and environments. In dilute
atomic gases~\cite{Young, B. Rudek}, for example, one should include
photoexcitations and ionizations, Auger and spontaneous decay
processes, and their reverse processes. For dilute
molecules~\cite{M. Hoener}, additional processes, such as photon
dissociations, should be taken into account. For solid-dense warm
and hot matters~\cite{S. M. Vinko}, electron impact excitations and
ionizations play an important role, in addition to environmental
screening effects~\cite{Li_2008}. Therefore, we develop a thousand
state master equation for describing {\it real} dynamics of these
complex systems induced by an X-ray laser, and discuss possible
experimental implementations such as signatures for coherent
evolution of inner-shell electrons. Our studies will provide the
basis for understanding the coherence effects in X-ray absorption
mechanisms at a fundamental level. Actually, if all the off diagonal
elements of the density matrix are ignored, our method reduces to
Einstein's rate equation approach, which is a semiclassical
description for X-ray-matter interactions~\cite{N. Rohringer2, O.
Ciricosta, W.-J. Xiang}. The present approach can be applied to many
different practical open quantum systems, which are widely treated
qualitatively before by a few-level master equation
approach~\cite{H.-P. Breuer}, such as those in atomic
physics~\cite{B. R. Mollow}, quantum optics~\cite{R. Brewer} and
cold matter physics~\cite{P. Zoller, J. Cirac}.

\begin{figure}[h!]
\vspace{-4mm}
\hspace{3mm}
\includegraphics[width=0.95\linewidth]{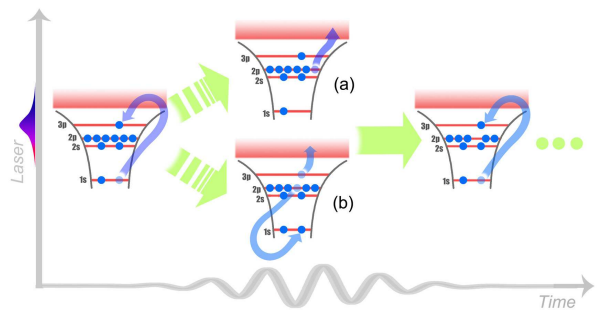}
\vspace{-7mm}
\caption{Sketch of multiphoton absorptions in neon
induced by ultra-intense X-ray pulses. Excitations and ionizations from the 1$s$ shell dominate the photoabsorptions, followed by the
consequent further outer-shell ionizations (a) and Auger decay processes (b), ending up with a highly ionized stage. Coherent Rabi oscillations interplay with dissipations in the X-ray-atom systems, such as Auger and spontaneous decay processes.}\label{sketch}
\end{figure}
In the recently typical experiment~\cite{Young}, however, atomic
gases are dilute and the influence of electron collisions and photon
scattering can be neglected~\cite{O. Ciricosta}. In this paper, we
take dilute atomic gases as examples for discussing coherent
dynamics of the rapidly decayed X-ray-matter systems (sketch in Fig.~\ref{sketch}), where there
are still few studies related to {\it real} dynamics of complex
atoms based on multilevel master equation approach~\cite{S. M.
Cavaletto}. Here, comparisons between master equation and rate
equation approach will be made to investigate the influence of
coherence on the dynamical mechanics, as related to the ongoing
experiments with different temporal coherence. In parallel, we also
perform an approximate calculation based on a degenerate atomic
master equation approach which is computationally more affordable.

The paper is organized as follows: in section II we give a detailed
description of the approach. Section III covers our results for
coherent dynamics of neon induced by an X-ray laser, and discusses
Auger processes of neon based on thousand state master equation
approach. We summarize with a discussion in Section IV.


\section{Theoretical Model} \label{sec:model}
\subsection{Hamiltonian}
We consider a many-body system, such as dilute atomic and molecular
gases~\cite{Young, B. Rudek, M. Hoener}, and solid-state
materials~\cite{S. M. Vinko}, coupled with incoherent sources such
as vacuum, irradiated by a high intensity X-ray laser. Inner-shell
electrons of these atoms and molecules will be excited, forming a
far-off-equilibrium system and typically relaxing in a fs timescale
via spontaneous, Coulombic and Auger decay processes. These
processes compete with other mechanics, such as processes including
coherent photoexcitations and ionizations. Correspondingly, the
total Hamiltonian of the X-ray-matter systems can be written as
\begin{eqnarray} \label{Hamil}
 \hat{H}=\hat{H}_A+\hat{H}_F+\hat{H}_I + \hat{H}_{\rm inc},
\end{eqnarray}
where the total Hamiltonian is the sum of the Hamiltonian
$\hat{H}_A$ of the many-body system in vacuum, the external field
Hamiltonian $\hat{H}_F$ including the coherent and incoherent
external field, the laser-matter interaction $\hat{H}_I$, and the
incoherent-field-matter interaction $\hat{H}_{\rm inc}$.

The Hamiltonian
\begin{eqnarray}
\hat{H}_A=\sum_{k=1}^{N}\epsilon_k \hat{A}_{kk},
\end{eqnarray}
governs the time evolution of the system in vacuum, such as dilute
atomic and molecular gases, and condensed materials, with
$\hat{A}_{kk} = |k\rangle\langle k|$ and $N$ being the total energy
levels included. Here, $|k\rangle$ and $\epsilon_k$ denote the
eigenstate and eigenvalue of the system, respectively. The
Hamiltonian of the external field is given by
\begin{eqnarray}\label{field}
\hat{H}_F = \sum_i\hbar \omega_i^a \hat{a}_i^\dagger \hat{a}_i + \sum_i\hbar \omega_i^b \hat{b}_i^\dagger \hat{b}_i,
\end{eqnarray}
where $\hbar$ denotes the reduced Planck constant, $\hat{a}_i$ ($\hat{a}_i^\dagger$) denotes the annihilation
(creation) operator that corresponds to the $i$-th mode of the laser
with frequency $\omega_i^a$, and $\hat{b}_i$ and $\omega_i^b$
denote those for the incoherent field.

With the semi-classical treatment for the laser field, the
quantization of the light is ignored. Instead the light is
considered as a electric field, ${\it E(t)}$, which interacts with the
$i$-th dipole ${\rm d}_i$ for the transition between states $|k\rangle$
and $|k^\prime\rangle$ to give
\begin{eqnarray} \label{Hamil_interaction}
\hat{H}_I=-\sum_{i}\frac{\hbar\Omega_i(t)}{2}(\hat{D}_ie^{-i\omega_Lt}+{\rm H.c.} ),
\end{eqnarray}
where the transition operator $\hat{D}_i=|k\rangle\langle
k^\prime|$, and Rabi frequency $\hbar\Omega_i(t)=e {\rm d}_iE(t)$ at time $\it t$
with ${\rm d}_i=\langle k^\prime |{\hat {\rm d}}_i | k \rangle$. We
remark here that we have used the dipole approximation for the
coupling in Eq.~(\ref{Hamil_interaction}), and can also trace out
the external field degrees of freedom within the semi-classical
treatment indicating the Eq.~(\ref{field}) can be ignored in the
simulations. This semiclassical approximation forms the basis of
most investigations on the many-body systems both coherently and
incoherently coupled by a strong external field.

For magnetic sublevels, we rewrite the Hamiltonian of the
matter-field interaction in a more explicit form
\begin{eqnarray}
H_I =\sum_{J,J^\prime}\frac{\hbar\Omega_i(t)}{2}({D_{JJ^\prime\sigma}e^{-i\omega t}} + {\rm H.c.}),
\end{eqnarray}
where $\hbar\Omega_i(t)=eE(t)\langle J |{\hat {\rm d}}_i | J^\prime \rangle$ and
\begin{eqnarray}
D_{JJ^\prime \sigma} =(-1)^{J-M_J}
                              \left( \begin{array}{ccc}
                               J   &  1   & J^\prime \\
                               -M_J &  \sigma   & M_J^\prime
                              \end{array} \right) |J,m_J\rangle \langle J^\prime, m_{J^\prime}|. \nonumber
\end{eqnarray}
Here $\sigma$ denotes the polarization of the external laser, and
the dipole operator $\hat {\rm d}_i$ describes a $J \rightarrow J^\prime$
transition between states $|J, M_J\rangle$ and $|J^\prime,
M_J^\prime\rangle$ with Zeeman substructures.

Actually, we can remove the explicit time dependence from the Hamiltonian, if
Rabi frequency $\Omega $ is time independent and $\Omega \ll
\epsilon_i, \omega$, transforming the Hamiltonian of the system to
an arbitrarily specified rotating frame. The interaction Hamiltonian
in a frame rotating at the laser frequency $\omega$ reads
\begin{eqnarray}
H_I &=&\sum_{J,J^\prime}\frac{\hbar\Omega}{2}(D_{JJ^\prime \sigma}+ {\rm H.c.}).
\end{eqnarray}

The system can also be pumped by an incoherent field, such as the
black-body radiation field in a hot plasma environment, and it reads
\begin{eqnarray}
{\hat H}_{\rm inc} = \hbar\sum_i\sum_n({\rm d}_i{\bf e}_i\cdot {\bf e}_n {\hat b}_n\hat{D}_i + {\rm H.c.}),
\end{eqnarray}
where ${\bf e}_i$ and ${\bf e}_n$ denote the directions of the dipole moment ${\rm d}_i$, the polarization of the incoherent field, respectively. Actually, the contributions of the incoherent field can directly be included in
the master equation approach~\cite{A. Imamoglu, X.-M. Hu}, as
discussed in Sec.~\ref{sec:method}.


\subsection{Method}\label{sec:method}
Now, the next question is how to theoretically simulate the time
evolution of complex systems in the presence of the X-ray laser,
described by Eq.~(\ref{Hamil}). In contrast to low-Z atomic and
molecular gases~\cite{Kenneth C. Kulander, M. Lewenstein, M.
Protopapas, Z.-X. Zhao, J. Zhao_2008, B. Zhang}, the X-ray-matter
system is dominated by ultrafast decayed mechanics in the
experimental timescales, such as spontaneous decay processes, where
one loses the possibility to keep track of the couplings between the
system and the environment with infinite degrees of freedom. In
solid-state materials, electron-ion, electron-electron and ion-ion
collisions occur rapidly and randomly, where screening and
broadening effects due to the solid-density environment should be
taken into account in dynamical simulations. Hence, a statistical
description for the X-ray-matter system is needed, and here we study
the time evolution based on a generalized thousand state master
equation approach for the reduced density matrix of the system,
where the degrees of freedom of both the environment and X-ray laser
have been traced out in a perturbative treatment. The generalized
thousand-level master equation for a complex system, coupled to
vacuum modes of the electromagnetic field and irradiated by an X-ray
laser and incoherent radiation field, reads
\begin{eqnarray}\label{master-eq}
\frac{d\hat{\rho}(t)}{dt} = -\frac{i}{\hbar}[\hat{H}, \hat{\rho}(t)] + \mathcal{L}\hat{\rho}(t),
\end{eqnarray}
where $\hat{\rho}=\sum_k p_k |k\rangle \langle k |$ denotes the
reduced density matrix operator of the multilevel system, and
$\mathcal{L}\hat{\rho}(t)=\sum_i\Gamma_i/2[2\hat{D}_i
\hat{\rho}(t)\hat{D}_i^\dagger - \hat{D}_i^\dagger
\hat{D}_i\hat{\rho}(t) - \hat{\rho}(t) \hat{D}_i^\dagger \hat{D}_i]$
with $\Gamma_i$ denoting transition rate for the $i$-th dipole due to the background
radiation pump, spontaneous, Coulombic and Auger decays,
photodissociations, and collision processes. Here, the first term in the right side describes the
coherent dynamics of the multilevel system coupled with the laser
field, and the second term denotes the incoherent processes, such as
spontaneous and Auger decays, transitions due to the black-body
radiation field and collision processes. The contributions of the
black-body radiation pump are only nontrivial in warm and hot
matters, and collision processes can be neglected for the dilute
atomic and molecular gases since it occurs in a fs timescale being
much shorter than the average particle-collision time. Note that the
broadening contributions of the incoherent-field-system interactions
in Eq.~(\ref{Hamil}) and plasma environments are included to the
second term in Eq.~(\ref{master-eq}), while the corresponding energy
shifts are incorporated in the energy levels of the system in
vacuum. We remark here that the thousand state master equation has
unique features, which needs large-scale simulations for propagating
in time a matrix of $\approx 10^6\times10^6$ and is not trivial to
solve directly, such as stability of numerical linear algebra and
parallel procedure demanding.

There are two possible ways to take photoionization processes into
account in our simulations. While we treat the ionization as
incoherent processes by adding photoionization cross section in the
incoherent terms for the low intensity X-ray laser, we consider
ionization as coherent processes for multiphoton dominant
processes, which features analogies to bound-state transitions of the system. For
multiphoton dominant processes, the ionized state composing
of the residual system and ionized electrons reads
\begin{eqnarray}
|k\rangle = | j_{\rm core}, \kappa; J,\epsilon_k,P\rangle,
\end{eqnarray}
where $j_{\rm core}$, $\kappa$, J, $\epsilon_k$, P denote the
angular momentum of the residual system, relativistic angular momentum
of free electrons, total angular momentum, total energy and parity
of the system, respectively. In the physical systems, these states can be
populated by multiphoton excitations. Considering the large amount
of continuous states, selection rules should be used to solely
include dominant states, such as degenerate initial, intermediate
and finial continuous states connected by multiphoton energies, and
at the same time neglect states detuned from resonance excitations
since the finite time duration of the X-ray laser implies
a finite Fourier width for the bound-free or free-free transitions.
These selections are a good approximation for hydrogen in the
presence of strong laser beams, and therefore we anticipate that
the dominant states for ionizations of a complex system should be
similar.

Complementary to the multilevel master equation, in this work
we employ a degenerate master equation approach to explore the
physics of Eq.~(\ref{Hamil}), which is computationally more
affordable. Here, the Rabi frequency is $\sqrt{(2M_J+1)(2M_J^\prime+1)}\Omega_i$ and the decay
rate is defined as a total transition probability from one upper
state $|J, M_J\rangle$ to all the degenerate lower states of the level
$|J^\prime\rangle$.

\paragraph{population distribution---}
Free states $|k\rangle$ form a complete basis set, and we can
work in the set and obtain a number of algebraic equations for
Eq.~(\ref{master-eq}). Due to infinite number of matrix elements,
however, truncation for a finite $N$-state basis is required. And
then the full Eq.~(\ref{Hamil}) is projected onto this subspace
spanned by the $N$-basis states, and it is expected this projection
gives the best possible description of dynamics of the system
induced by an X-ray laser, such as a neon gas by including thousands
of energy levels.
After solving the multilevel master equation, we obtain population
distributions for each level and coherence between different
transition states as a function of time, by calculating
$\rho_{kk^\prime}=\langle k| \hat \rho(t) | k^\prime \rangle$, and
then comparison can be made with experimental data by utilizing ion
charge-state spectra recorded by a time-of-flight
analyzer~\cite{Young}, even though the recently typical
free-electron lasers have limited temporal coherence and new
techniques are needed to improve it.

Master equation approach can reduce to Einstein's rate equation
approach, if the pump field is fully stochastic, such as a broadband
isotropic light field. In this case, coherence embedded in the
off-diagonal terms is neglected, and the time evolution of the
system is characterized by the population changes for each energy
level.

\paragraph{resonance fluorescence---}
The spectrum of resonance fluorescence in high-intensity x-ray
pulses can also be calculated via master equation approach~\cite{M.
Lax, J. H. Eberly, M. Florjanczyk, M. Wilkens, P. Zoller_1993, S. M.
Cavaletto}. The time-dependent spectrum of the fluorescent light,
described by autocorrelation function of the electric field operator
$\hat{E}(t)$, is given by~\cite{R. J. Glauber, P. Zoller_1993}
\begin{eqnarray}
S(t,\omega)=\int_{0}^t\int_{0}^t e^{-i\omega(t_1-t_2)}\langle \hat{E}^-(t_1) \hat{E}^+(t_2)\rangle dt_1dt_2,
\end{eqnarray}
where $\hat{E}^-(t)$ and $\hat{E}^+(t)$ denote negative and
positive frequency parts of the electric field, respectively.

The change in time of the occupation number of photons can be
related to the dipole moment operator, and it reads
\begin{eqnarray}
\hat{E}^{-}(t) = C({\bf r}) \hat{D}_i^\dagger(t),
\end{eqnarray}
where $C({\bf r})$ is a proportionality factor at position ${\bf r}$
and can be neglected in a homogeneous system. And then we obtain
\begin{eqnarray}
S(t,\omega)=2\int_{0}^t dt_2 \int_{0}^{t-t_2} {\rm Re}\bigg[e^{-i\omega\tau}\langle \hat{D}_i^\dagger(t_2+\tau)\hat{D}_i(t_2) \rangle \bigg] d\tau,
\end{eqnarray}
where we only include the contributions in the region $t_1\geq t_2$
and introduce the time delay $\tau\equiv t_1-t_2$.

Thus we need the knowledge of two-time expectation values of
$\hat{A}_{ij}(t_1,t_2) = \hat{D}_j(t_1) \hat{D}_i(t_2)$ for the
time-dependent fluorescent spectrum of the $i$-th dipole transition.
Applying the quantum regression theorem~\cite{M. Lax, J. H. Eberly},
we obtain the coupled equation for the $i$-th dipole transition,
\begin{eqnarray}
\frac{d\hat{A}(t_1,t_2)}{dt_1} = -\frac{i}{\hbar}[\hat{H}(t_1), \hat{A}(t_1,t_2)] + \mathcal{L}\hat{A}(t_1,t_2),
\end{eqnarray}
where $t_1\geq t_2$, and the initial condition can be obtained by
the one-time evolution operator $A_{ij}(t_2,t_2) = \langle
\hat{D}_j(t_2) \hat{D}_i(t_2)\rangle$ via Eq.~(\ref{master-eq}).

\subsection{Examples: dilute atomic gases}
Before proceed to construct the master equation, we should first
calculate the wavefunctions and energy levels of the system, and
obtain the required data, including oscillator strength, dipole
moment, Rabi frequency, spontaneous and Auger decay rates, and
photoionization cross section. Actually, dilute atomic gases are
widely used in the recently typical experiment~\cite{Young}. From
now on, we take dilute multielectron atomic gases as examples for
discussing coherent dynamics of the rapidly decayed X-ray-matter
systems. Without loss of generality, these discussions can be easily
applied to other X-ray-matter systems, such as molecular gases and
solid-state materials by calculating corresponding wavefunctions,
energy levels and microscopic transition processes. Here, we can map
the dilute atomic system into a single-atom problem each of which
can be written as in Eq.~(\ref{Hamil}). The computations of various
atomic radiative processes involve the bound and continuum states
of different successive ionization stages in the single atom. For
the bound states, a fully relativistic approach based on the Dirac
equation is utilized, while for the continuum processes the
distorted wave approximation is employed. The bound states of the
atomic system are calculated in the configuration interaction
approximation. The radial orbitals for the construction of basis
states are derived from a modified self-consistent Dirac-Fock-Slater
iteration on a fictitious mean configuration with fractional
occupation numbers, representing the average electron cloud of the
configurations included in the calculation. The detailed discussion
can be found in Refs.~\cite{Gao_2013, Li_2008, Gao_2}.

\paragraph{Rabi frequency---}
In the dipole approximation, the Rabi frequency for the bound-state
transition can be obtained from the degenerate emission oscillator
strength of the electric dipole ${\hat{\rm d}}_i$,
\begin{eqnarray}
gf_i &=& \frac{2m_e}{3\hbar^2}\Delta E\, |\langle \hat{\rm d}_i \rangle|^2 \nonumber \\
   &=& 8.749\times10^{18} \times \Delta E \, |\langle \hat{\rm d}_i \rangle|^2,
\end{eqnarray}
and then it yields.
\begin{eqnarray}\label{rabi_gf}
\Omega_i &=& \frac{eE}{\hbar}\langle \hat{\rm d}_i \rangle \nonumber  \\
       &=& 1.409\times10^9 \sqrt{I\times gf_i/\Delta E } \nonumber \\
       &\times&(-1)^{J-M_j}\left( \begin{array}{ccc}
                               J   &  1   & J^\prime \\
                               -M_J &  q   & M_J^\prime
                              \end{array} \right),
\end{eqnarray}
where laser intensity $I$ and transition energy $\Delta E$ are in
units of ${\rm W/cm^2}$ and ${\rm eV}$, respectively. Here $q$
denotes the polarization of the external laser, ${\rm {\hat d}}_i$
denotes the electric dipole moment for the $J \rightarrow J^\prime$
transition between states $|J, M_J\rangle$ and $|J^\prime,
M_J^\prime\rangle$ with Zeeman structure, $m_e$ denotes the mass of
atom, E denotes the electric field, and $e$ denotes the charge of
electron.

For photoionization processes (bound-free transition), the cross
section is given by the differential oscillator strength
$\frac{df}{d\epsilon}$,
\begin{eqnarray}
\sigma(\epsilon) &=& 4\pi^2\alpha a^2_0 \frac{df}{{d}\epsilon} \nonumber \\
       &=& 8.067\times10^{-18} \times \frac{df}{d\epsilon},
\end{eqnarray}
where the energy $\epsilon$ is in unit of Ry, $\alpha$ denotes the
fine-structure constant and $a_0$ denotes the Bohr radius. And then
Rabi oscillation for bound-free processes can be written as:
\begin{eqnarray}
\Omega &=& \frac{eE}{\hbar}\langle \hat{\rm d}_i \rangle \\
       &=& 0.135 \times 10^{18} \int g(\epsilon)\sqrt{I \times \sigma(\epsilon)  /\Delta E }\,d\epsilon,  \nonumber
\end{eqnarray}
where, $I$, $\sigma$ and $\Delta E$ are in units of ${\rm W/cm^2}$,
${\rm cm^2}$ and ${\rm eV}$, respectively, and $g(\epsilon)$ is the
lineshape of the bound-free transition, such as broadening due to
the finite period of time duration for the X-ray laser.

For free-free processes, the Rabi frequency of the
$\kappa\rightarrow\kappa^\prime$ transition with the residual ion in
the state $|j_{\rm core}\rangle$ can be given by:
\begin{eqnarray}
\Omega = \frac{eE}{\hbar}\langle j_{\rm core}, \kappa; J | \hat{\rm d}_i |  j_{\rm core}, \kappa^\prime; J^\prime \rangle,
\end{eqnarray}
where $\kappa$ and $J$ denote the relativistic angular momentum of
free electrons and the total angular momentum of the system,
respectively.

\paragraph{Spontaneous decay---}
Even in the absence of an applied field, spontaneous emission cannot
be ignored, since the excited state interacts with the vacuum
fluctuations of the electromagnetic field. After integrating over
all possible modes and summing over the two orthogonal polarizations
possible for each wave vector, the spontaneous $|e\rangle
\rightarrow |g\rangle$ transition rate is given by a perturbation
treatment,
\begin{eqnarray}
A_{i}  &=& \int g(\omega)|\Omega_{i}|^2 d\omega \nonumber  \\
       &=& \frac{e^2\omega^3_0}{3\pi\epsilon_0\hbar c^3} |\langle g| \hat {\rm r}|e\rangle |^2,
\end{eqnarray}
where $c$ is the speed of light, $\epsilon_0$ denotes the
permittivity, and $\omega_0$ denotes the transition frequency. Here,
one assumes the lineshape function $g(\omega)$ is sharply peaked
around $\omega_0$, so that $\int_\omega  g(\omega) |\langle g| \hat
{\rm r}|e\rangle |^2 \omega^3  d^3\omega \approx |\langle g| \hat
{\rm r}|e\rangle |^2 \omega^3_0   \int_\omega  g(\omega) d^3\omega =
2\pi \omega^3_0 |\langle g| \hat {\rm  r}|e\rangle |^2$.

\paragraph{Stimulated transition---}
If the $|g\rangle \rightarrow |e\rangle$ transition is driven by the
external radiation field with photon polarization $\epsilon$ in the
differential solid angle $d\Theta$, the transition rate in this
solid angle $\Theta$, based on Fermi's Golden rule, is written as:
\begin{eqnarray}
dB_{i} &=& \int \frac{e^2}{\hbar^2} |\langle e| \hat {\bf r}\cdot \epsilon |g\rangle |^2 \frac{\hbar\omega}{2\epsilon_0V} \bigg(n^\epsilon_\lambda \frac{V}{(2\pi c)^3} \omega^2\bigg) g(\omega) d\omega d\Theta \nonumber \\
&=& \int \frac{e^2\omega^3}{2\hbar\epsilon_0 (2\pi c)^3} |\langle e| \hat {\bf r}\cdot \epsilon |g\rangle |^2 n^\epsilon_\lambda \,  g(\omega) d\omega  d\Theta
\end{eqnarray}
with $|E|^2 = \frac{\hbar\omega}{2\epsilon_0V}\, n^\epsilon_\lambda \frac{V}{(2\pi c)^3} \omega^2$.

As for black-body radiation field, energy density of its radiation varies
slowly over the range of transition energies,
\begin{eqnarray}
\int_\omega g(\omega) \rho(\omega) d\omega \approx  \rho(\omega) \int_\omega g(\omega)  d\omega = 2\pi \rho(\omega).
\end{eqnarray}
Considering the isotropic and unpolarized properties of the
radiation field, the total transition rate can be obtained by
integrating over all possible modes and summing over the two
polarizations of the field:
\begin{eqnarray}
B_{i} =  \frac{e^2\omega^3_0}{3\pi\epsilon_0\hbar c^3} n_\lambda |\langle e| \hat {\bf r}|g\rangle |^2,
\end{eqnarray}
and correspondingly Einstein B coefficient
\begin{eqnarray}
B_{Ein} =  \frac{\pi e^2}{3\epsilon_0\hbar^2}|\langle e| \hat {\bf r}|g\rangle |^2,
\end{eqnarray}
with energy density $\rho(\omega) = \hbar \omega^3 n_\lambda/\pi^2c^3$. We remark here that the transition rate $B_i=A_{i}n(\omega)$, with
$n(\omega)$ being the mean photon number of the incoherent field at
frequency $\omega$~\cite{A. Imamoglu, X.-M. Hu}, such as the black-body radiation field.

As for laser driven atomic system, one can use the relation between
laser intensity $I$ and photon number $n^\epsilon_\lambda$ with
polarization $\epsilon$ and wavelength $\lambda$
\begin{eqnarray}
I(\Theta)/c = n_\lambda \frac {\omega^2}{(2\pi c )^3} \hbar \omega,
\end{eqnarray}
and then obtains:
\begin{eqnarray}
\frac{dB_{i}}{d\Theta} &=& \int \frac{e^2}{2\hbar^2 \epsilon_0 c } |\langle e| \hat {\bf r}\cdot \epsilon |g\rangle |^2 I(\Theta)  g(\omega) d\omega  \nonumber \\
                        &=& \int \frac{ e^2}{6\hbar^2\epsilon_0 c} |\langle e| \hat {\bf r} |g\rangle |^2 I(\Theta)  g(\omega) d\omega ,
\end{eqnarray}
where one averages over all the polarization directions of the dipole
moment. Normally, the laser injects from one certain direction with
central frequency $\nu$
($I(\Theta)d\Theta=I(\nu)\delta(\Theta)d\Theta $), and in this case
the total transition rate can be written as:
\begin{eqnarray}
B_{ge} = \frac{\pi e^2}{3\hbar^2 \epsilon_0 c} |\langle e| \hat {\bf r} |g\rangle |^2 \int I(\nu) g(\nu) d\nu
\end{eqnarray}
where $g(\nu)$ is the line shape of the $|g\rangle \rightarrow |e\rangle$ transition.

If one defines absorbtion (emission) cross section as:
\begin{eqnarray}
\sigma_{i}(\nu) &=& B_{i} h\nu/I(\nu) \nonumber \\
                 &=& \frac{2\pi^2 e^2\, \nu}{3\epsilon_0\hbar c}  \Big|\langle e| \hat {\bf r} |g\rangle \Big|^2 g(\nu),
\end{eqnarray}
one obtains
\begin{eqnarray}
B_{i} &=& \int_\nu \sigma_{i}(\nu) \frac{I(\nu)}{h\nu}d\nu \\
       &=& \int_\nu \frac{I(\nu)}{h\nu} \frac{\pi e^2}{2\epsilon_0 m_e c} f_{i} g(\nu) d\nu \nonumber \\
       &=&  109.761\times 10^{-18}\times  \nonumber \\
        &&\int_\omega \frac{I(\omega)f_{i}}{\hbar\omega} \frac{\hbar A_{i}/2\pi}{(\hbar\omega-\hbar\omega_{i})^2 + (\hbar A_{i}/2)^2}d\omega, \nonumber
\end{eqnarray}
where the full-width-half-maximum $\Delta\nu=A_{i}/2\pi$, and laser
intensity $I$ and photon energy $\hbar \omega$ are in units of ${\rm
W/cm^2}$ and ${\rm eV}$, respectively.

\paragraph{Autoionization---}

Complex atoms, irradiated by X-ray laser, can form hole atoms and
then relax in a fs timescale via Auger decay processes, i.e.
spontaneously emitting one of the electrons in along with
refilling the hole by outer-shell electrons. The corresponding autoionization rate is expressed as
\begin{eqnarray}
A_i^a = 2\sum_\kappa \Bigg|\bigg\langle j_{\rm core},\kappa; J, M_J\bigg|\sum_{m<n}\frac{1}{{\bf r}_{mn}}\bigg|J^\prime, M_J^\prime \bigg\rangle\Bigg|^2,
\end{eqnarray}
where $\kappa$ is the relativistic angular quantum number of the
ionized electron, and ${\bf r}_{mn}$ denotes the relative
displacement of between electrons $m$ and $n$. Here, $J$ and $M_J$
denote total angular moment and magnetic quantum number of the
system consisting of the residual ion and ionized electrons,
respectively.

For some cases, double Auger processes are nontrivial and should be
included in the simulations, and the code for the atomic data of
direct two-electron processes has been developed recently~\cite{J.
Zeng}. In this work, we do not include the contributions from the
direct double Auger processes, yet its implementation in the master
equation is straightforward.

\FloatBarrier
\section{Results}
In this section, we discuss coherent dynamics of the rapidly decayed
X-ray-matter systems. As examples, we investigate ultrafast dynamics
of inner-shell electrons of complex atoms irradiated by an X-ray
laser via a thousand state master equation approach. The influence
of coherence on dynamical mechanics will be investigated through
comparisons between atomic systems induced by X-ray pulses with
different temporal coherence. For ionizations, multiphoton ejections
of electrons from an atom subjected to a strong laser field will be
investigated in the framework of master equation approach, where the
reliability of our numerical results will be verified against the
time-dependent Schr\"odinger equation. For inner-shell dynamical
processes, Auger and spontaneous decay processes typically occur in a
fs timescale and compete with other mechanics, such as
coherent Rabi oscillations, where we will discuss possible
experimental implementations such as signatures for coherent
evolution of inner-shell electrons. Finally, we will discuss a {\it
real} coherent dynamics of complex atoms induced by an X-ray laser,
based on master equation approach by including thousands of atomic
levels.

\subsection{Coherent dynamics between bound states}
\begin{figure}[t]
\begin{tabular}{c}
\includegraphics[width=\linewidth]{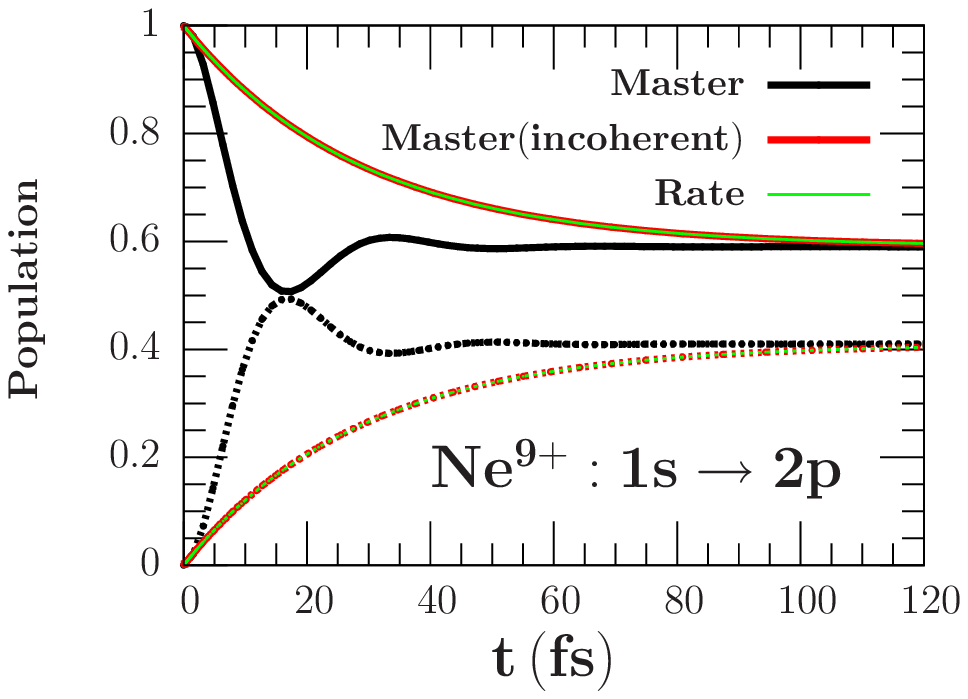}
\vspace{-3mm}
\\
\includegraphics[width=\linewidth]{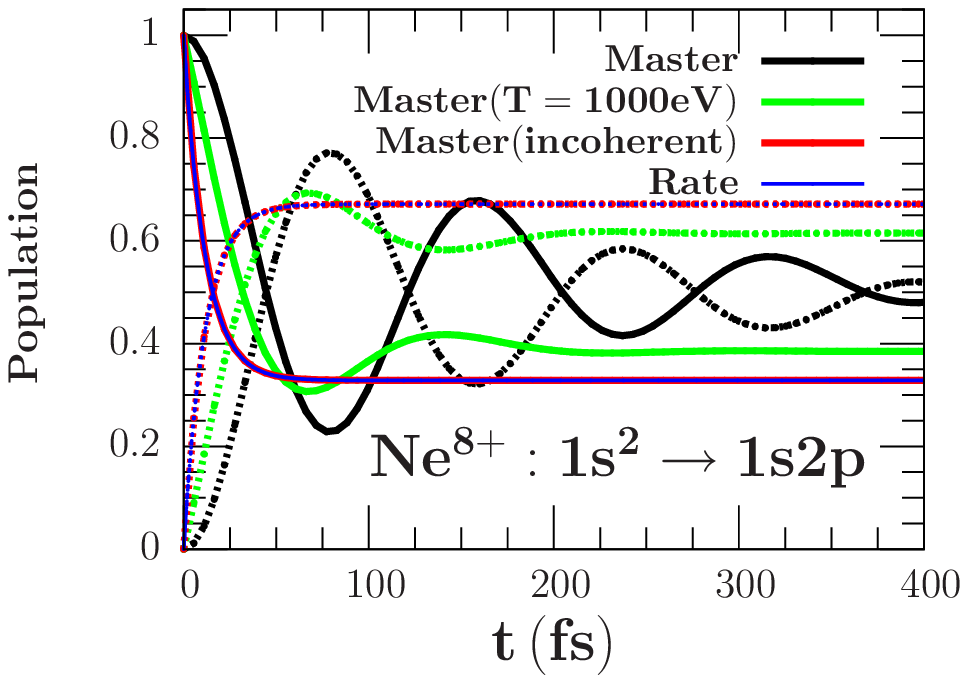}
\end{tabular}
\vspace{-3mm} \caption{Comparison with rate equation. {\bf Upper}:
State populations of a two-level H-like neon (between
$|J=0.5,M_J=0.5\rangle$ of $1s$ and $|J=1.5,M_J=1.5\rangle$ of $2p$)
as a function of time. {\bf Lower}: Degenerate state populations for the transition
$|J=0\rangle\rightarrow |J=1\rangle$ ($1s^2$-$1s2p$) of He-like neon. Here, the solid and dashed lines denote populations of the ground and excite states, respectively, coupled by a laser with an intensity $I_0=10^{12}\,{\rm W/cm^2}$. For the
incoherent pump, master equation yields identical results with those from the
rate equation.}\label{rate_master}
\end{figure}
In this section, we investigate coherent dynamics between bound
states of an atomic system coupled by a X-ray laser, and pay special
attention to the influence of coherence on dynamical evolution by
comparising with Einstein's rate equation, which is a reduction of
master equation approach and a method for the system
illuminated by a broadband isotropic light field (incoherent light).

First we study the time evolution of a two-level H-like neon with
atomic data being obtained via solving Dirac
equation~\cite{Gao_2013, Li_2008, Gao_2}. Due to the pump of the
laser beam and coupling with the external environment, it is
expected that the system exhibits a Rabi-oscillating feature, and
then decays to a steady state after a longer time. As shown in the
upper panel of Fig.~\ref{rate_master}, we observe Rabi oscillations
between $|J=0.5,M_J=0.5\rangle$ (1s) and $|J=1.5,M_J=1.5\rangle$
(2p), and, after a few fs, steady states are achieved with the
excited-state population being $\sqrt
{\Omega^2/(\Gamma^2+2\Omega^2)}$. We observe that the stable states
of master equation at a longer time are identical with rate equation
approach, whose conclusion is also consistent with analytical
results.

Then we investigate a multilevel atom interacting with an isotropic
and unpolarized radiation beam, as relevant to Einstein's rate
equation. We focus on the dipole transition $|J=0\rangle \rightarrow
|J=1\rangle$ ($1s^2$-$1s2p$) of He-like neon, where there are four
states in the reduced Hilbert space and the upper three levels with
angular momentum $|J=1\rangle$ are triply degenerate due to magnetic
splitting. We assume the ground state $|J=0\rangle$ are populated at
$t=0$, and then switch on the pump laser. We observe a decayed
Rabi-flopping structure in the master equation, due to the coupling with
the pump laser and the external environment, while in the rate
equation there is a monotonous decay for the ground state, as shown
in the lower panel of Fig.~\ref{rate_master}. We also observe that
both methods cannot yield identical stable states at a longer time.
It indicates that, for a multilevel atom coupled by a laser field,
coherence influences both short-time structures and long-time
stable states, and Einstein's rate equation approach cannot give
reasonable results since coherence is totally neglected in this
method. The coherence effects can also be clearly seen in Sec.~\ref{real}.

Another issue needed to be addressed is the influence of different temporal
coherence of the pump field on the dynamics of complex atoms. To
simplify our discussion, an incoherent thermal radiation field is
taken into account as incoherent pump sources, in addition to a
coherent X-ray laser. It is expected that the interplay between
coherent and incoherent pump influences dynamical properties of the
atomic system. For example, the incoherent thermal radiation field
can be introduced via $R_{J,J^\prime}=\Gamma_{J,J^\prime}\tilde{n} =
\Gamma_{J,J^\prime}\{{\rm exp}[\hbar (\omega_{J^\prime}-
\omega_J)/kT]-1\}$, where $T$ is the temperature of the incoherent
thermal radiation field, $\tilde{n}$ is average number of thermal
photons per mode at each transition frequency, and
$\Gamma_{J,J^\prime}$ denotes the spontaneous decay induced by
vacuum (the energy shift due to coupling with vacuum is included in
${\hat H}_a$). In the lower panel of Fig.~\ref{rate_master}, we show
results for an atomic system coupled with a coherent field and a
thermal radiation field of $T=1000$ eV, and find that the stable
states shift closer to those obtained from rate equation. For a
fully incoherent pump in the framework of master equation approach,
we find that the time evolution of the multilevel system coincides
with Einstein's rate equation approach.

\subsection{Coherent dynamics for photoionizations}
\paragraph{Comparison with time-dependent Schr\"odinger equation for hydrogen---}
\begin{figure}[h!]
\vspace{0mm}
\begin{tabular}{c}
\includegraphics[width=1\linewidth]{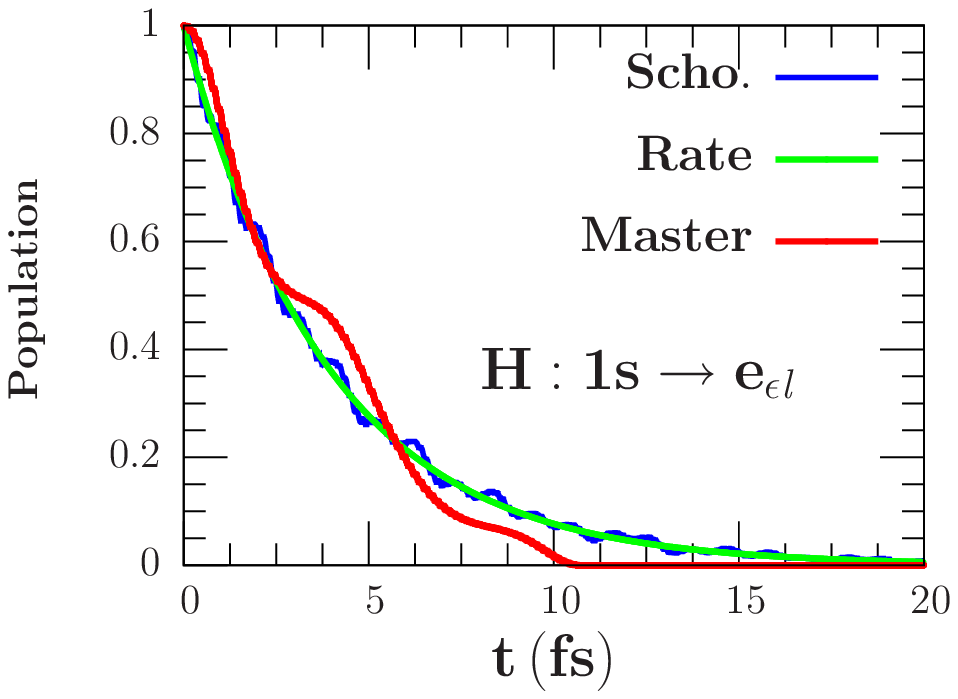}
\\
\includegraphics[width=1\linewidth]{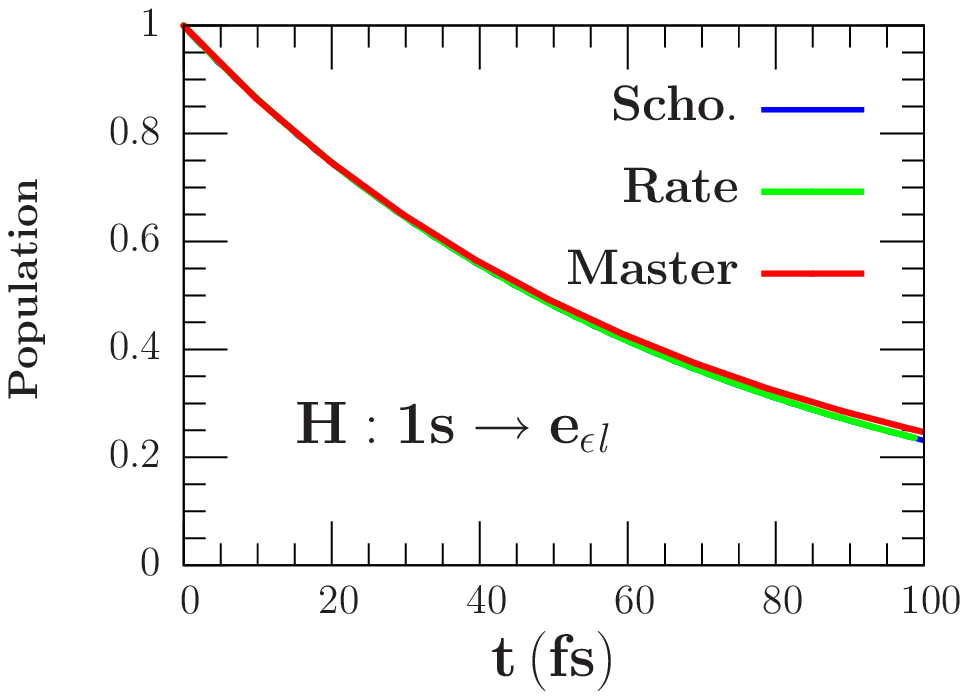}
\end{tabular}
\vspace{1mm} \caption{Comparison with time-dependent Schr\"odinger
equation. Ground-state populations of hydrogen as a function of
time for a laser intensity $I_0=10^{14}\, {\rm W/cm^2}$ with photon
energies of 14 eV (upper) and 30 eV (lower), respectively, obtained via
rate equation, master equation and time-dependent Schr\"odinger
equation.}\label{ionization_H1}
\end{figure}

Ionization is a basic process of atoms subjected to an external
electromagnetic field. For weak laser fields, perturbation theory
can be utilized to capture these processes. For example, one considers a system prepared
in an initial state $|i\rangle$ and perturbed by a periodic harmonic
potential $V(t) = Ve^{i\omega t}$ which is abruptly switched on at
time $t = 0$. Based on time-dependent perturbation theory, the first
two basis coefficient $c_n$ can be expanded as: $c^{(1)}_n =
\frac{2\pi}{\hbar^2}|\langle f |V| i \rangle| ^2
\delta(\omega_{fi}-\omega)$ corresponding to Fermi's golden rule,
and $c^{(2)}_n = \frac{2\pi}{\hbar^4}\sum_m |\frac{\langle f |V|
m\rangle \langle m | V|  i \rangle}{\omega_m - \omega_i - \omega } |
^2 \delta(\omega_{fi} - 2\omega)$ associating with two-photon
processes, where $|i\rangle$, $|m\rangle$ and $|f\rangle$ denote
the initial, intermediate and final states, respectively. In the
study of atomic dynamics, one can simply include photon
ionizations as decay coefficients up to the corresponding order,
just as used in rate equation approach, where the coherence in
ionization processes is totally neglected. For intense laser fields,
however, multiphoton ejections of one electron typically occur, and
manifest as dominant mechanics~\cite{S. Palaniyappan, W. Ackermann,
A. A. Sorokin, Young}. In this case, the perturbation theory lacks
the possibility for understanding multiphoton absorptions, and a
nonperturbative treatment is inevitable to address this issue. Here,
we describe coherent multiphoton ionizations in the framework of the
atomic master equation approach, and the validity of approach will
be verified via comparisons with time-dependent Schr\"odinger
equation where both the discreet and continuous states are treated
in the same level.

In the calculations here, remarkable agreements between master
equation and time dependent Schr\"odinger equation have been obtained
for real dynamics of hydrogen in the intermediate-wavelength
regime. In Fig.~\ref{ionization_H1}, for example, we show the
ground-state evolution of hydrogen induced by a strong laser
field with an intensity of $10^{14}$ ${\rm W/cm^2}$, and photon
energies of 14 eV and 30 eV, respectively, and find that the master
equation approach can capture photoionization processes of hydrogen
in the intermediate-wavelength regime. Surprisingly, rate equation
offers another possibility to address this issue at these parameter
regimes, even though the detailed structures are different, which
indicates that the leading order process dominates ionizations
here, irrespective of coherence or incoherence. We anticipate that
photoionization mechanism for complex atoms in the X-ray-wavelength regime
should also be captured by master equation approach.

\paragraph{Coherence in photoionizations of neon---}
In this section, we discuss ionizations of complex atoms in the short-wavelength regime. In this regime, time-dependent Schr\"odinger equation encounters difficulties for dealing with electron-electron correlations and irreversible processes, such as Auger and spontaneous decay processes, while rate equation approach, mainly based on photoionization cross sections, fully discards coherence effects in the dynamical processes. The underlying physics for ionizations of both outer- and inner-shell electrons of complex atoms and its relevant dynamics induced by intense X-ray lasers are still unclear. Here, we will address this issue related to ionizations of complex atoms driven by a strong X-ray laser, based on master equation approach whose validity has been verified against ionizations of hydrogen as a benchmark.

\begin{figure}[h!]
\begin{tabular}{c}
\includegraphics[width=1\linewidth]{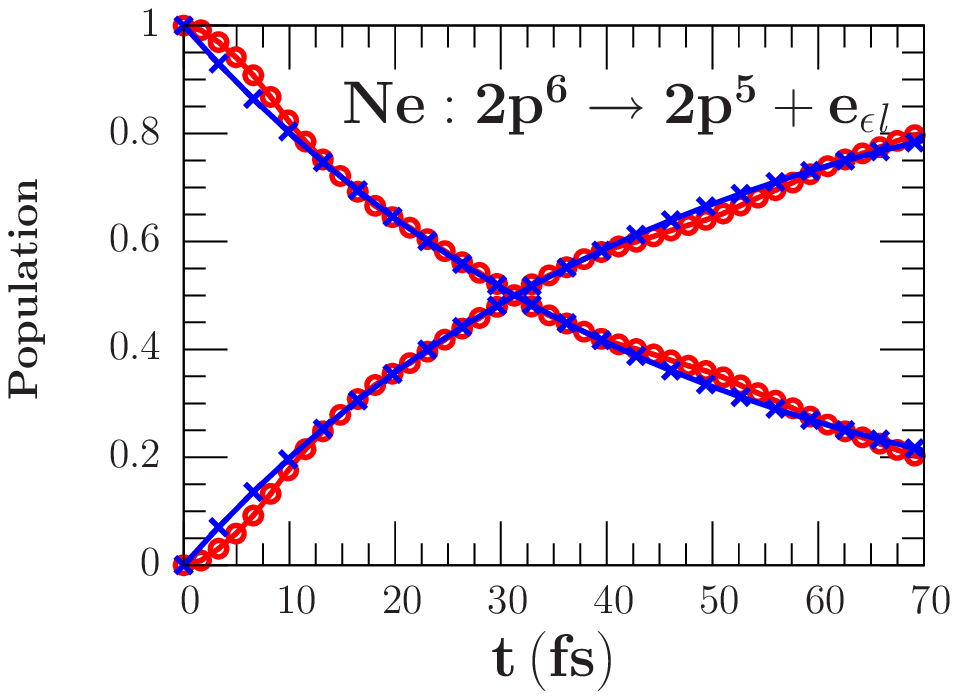}
\\
\includegraphics[width=1\linewidth]{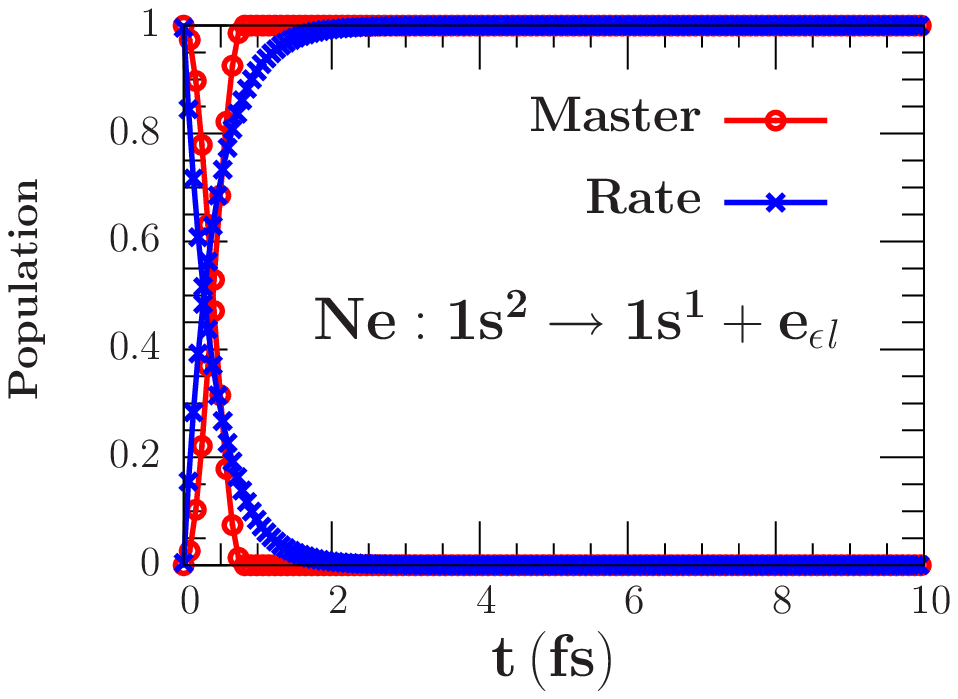}
\end{tabular}
\vspace{-2mm} \caption{Ionization of valence electrons (upper) and
inner-shell electrons (lower) of a neon gas induced by an X-ray
laser for a laser intensity of $10^{18}$ ${\rm W/cm^2}$ with photon
energy of 875 eV, obtained by master equation (red) and Einstein's
rate equation approach (blue), respectively. We find that
single-photon processes for both outer- and inner-shell electrons
dominate ionizations for the recently typical experiments with a
laser intensity of $\approx10^{18}$ ${\rm W/cm^2}$.}\label{Ne_4}
\end{figure}

In contrast to simple species such as hydrogen, the responses of neon
subjected to X-ray laser beams for valence and inner-shell electrons
are different, since Rabi frequency for valence electrons is
comparable to free-free excitations, while the frequency for
inner-shell electrons is much larger than free-free processes. We
find that inner-shell electrons are ionized much faster than
outer-shell electrons, as shown in Fig.~\ref{Ne_4} where ionizations
of valence (upper panel) and inner-shell electrons (lower
panel) of neon are triggered by an X-ray laser beam with a typical
experimental intensity of $10^{18}\, {\rm W/cm^2}$ and photon energy
of 875 eV. Comparisons between rate equation and master equation are
made and remarkable agreements are obtained, even though tiny
differences can be found in the detailed structures, due to coherent
oscillations for ionizations in master equation approach. As far as
we know, this is the first direct proof for coherent ionizations of
both outer- and inner-shell electrons of neon in the
short-wavelength regime, where sequential single-photon processes
dominate the absorptions for the recently typical experimental
conditions. This conclusion is consistent with the recent X-ray
experiments, where one did not observe clear signatures of
multiphoton single-electron ionizations. Hence it is not surprising
that comparisons between theories and experiments suggest the
dominant absorptions of electrons are sequential single-photon
processes in Young's experiment~\cite{Young}.

\subsection{Coherent dynamics in Auger decay processes}
\begin{figure*}
\vspace{-10mm}
\begin{tabular}{ccc}
\hspace{-3mm}
\includegraphics[width=0.435\linewidth]{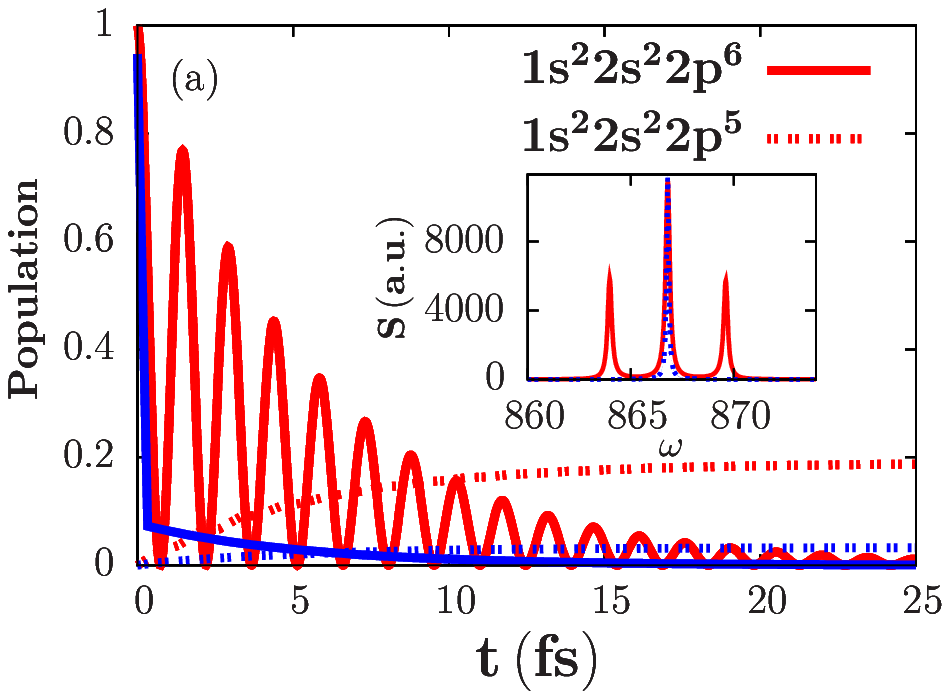}
\hspace{-20mm}
\includegraphics[width=0.435\linewidth]{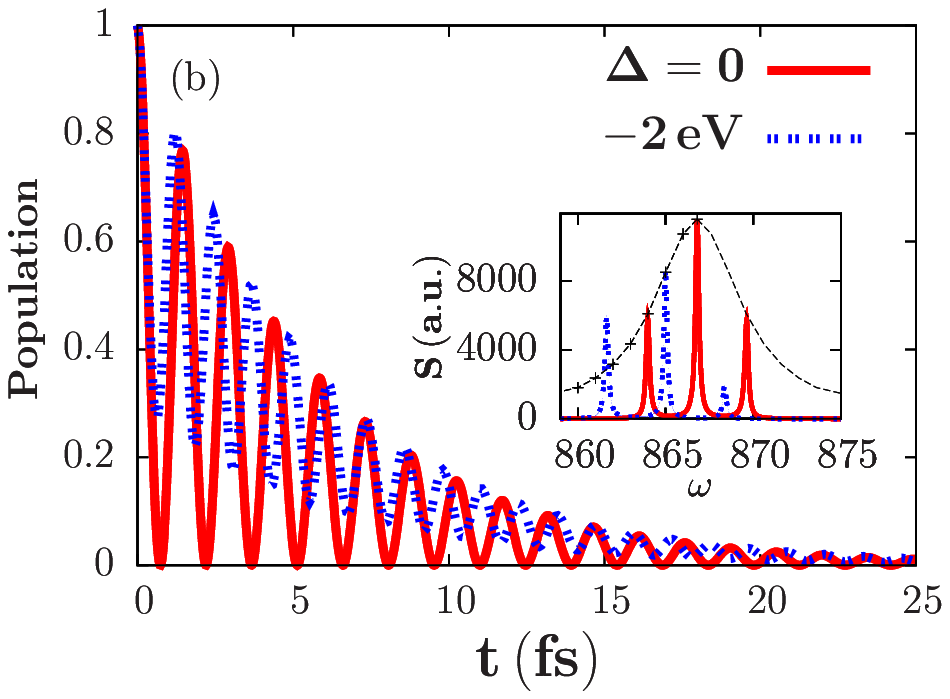}
\hspace{-20mm}
\includegraphics[width=0.435\linewidth]{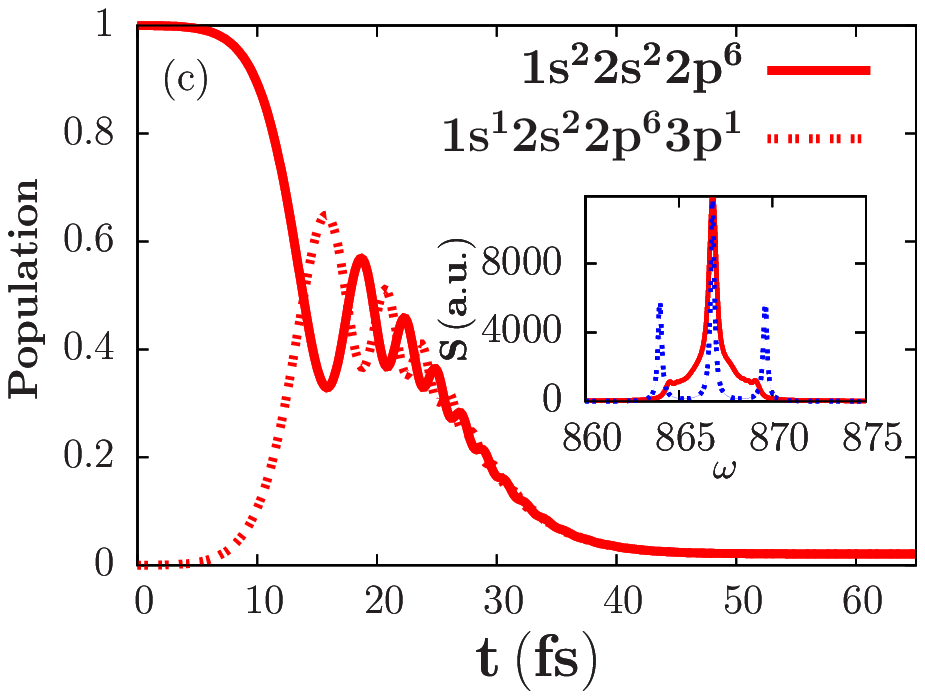}
\end{tabular}
\vspace{-5mm} \caption{Competition between Rabi oscillations, Auger
and spontaneous decays, and photoionizations of a neon gas
induced by an X-ray laser beam with a peak intensity of $10^{18}$
${\rm W/cm^2}$. (a) Populations of degenerate states $1s^22s^22p^6$
(solid) and $1s^22s^22p^5$ (dashed) coupled by a flat-topped X-ray
laser beam with a resonant photon energy for the $1s\rightarrow 3p$
transition, obtained via master equation (red) and rate equation
approach (blue), with the inset for the fluorescence spectrum of the
$1s^22s^22p^6\rightarrow 1s^12s^22p^63p^1$ transition. (b)
Ground-state populations of Ne for a resonant $1s\rightarrow 3p$
(red) and a red detuned pump $\Delta=-2$ eV (blue), where the inset
shows the fluorescence spectrum of the $1s^22s^22p^6\rightarrow
1s^12s^22p^63p^1$ transition with the black dashed line being the
power-broadening guideline for the central peak. (c) Neon gases
driven by a Gaussian X-ray pulse of a FWHM duration 15 fs, with the
inset for fluorescence spectrum of $1s^22s^22p^6\rightarrow
1s^12s^22p^63p^1$ transition (red) and those driven by a flat-topped
pulse with the identical fluence (blue).}\label{Ne_5}
\end{figure*}

In addition to ionizations of inner-shell electrons, Auger decay is another dominant process for complex
atoms, which carries valuable information about inner-shell electronic structures and
dynamical properties of atoms, molecules, and solids. Specifically,
inner-shell electrons can be excited or ionized by the X-ray laser
beam and a hole atom is formed. The atom is normally unstable and
relaxes through Auger and spontaneous decay processes, which
typically occurs in the fs timescale and can smear out
coherence-induced signals embedded in Rabi oscillations. In this
section, we will investigate the competitions between Rabi
oscillations, photoionizations, Auger and spontaneous decays, since
the magnitudes of these processes are typically of the same order. These
processes, however, are normally beyond the time revolution of the X-ray experiments. Fortunately, resonant fluorescence provides another
available tool for studying these competitions and underlying coherence effects.

In our simulations, we study dynamical evolution of ground-state
neon subjected to X-ray pulses with different temporal coherence
and resonant photon energy relative to $1s^22s^22p^6\rightarrow
1s^12s^22p^63p^1$ transition. After shining the X-ray laser on neon,
inner-shell electrons are coherently excited and coupled between
$1s$ and $3p$ orbitals. On the other hand, the hole state
$1s^12s^22p^63p^1$ is unstable and decays in a fs timescale, mainly
due to Auger decay processes by refilling the inner $1s$ orbital by
the outer-shell electrons, or spontaneous decays of $2p$ and $3p$ electrons. The hole state can also be further ionized with
double-core forming or via sequential valence electrons. To
investigate the interplay between these different processes driven
by X-ray pulses, we include all the dominant microscopic processes. We observe a decayed
Rabi-flopping structure between $1s$ and $3p$ states coupled by an
X-ray laser, based on master equation approach, while it is absent
in rate equation method with a monotonous population changing. It
indicates that the atomic coherence is distinctly embodied in Rabi
oscillations driven by the X-ray laser, in
spite of extremely fast Auger decays in the fs timescale.
At present, however, one lacks reliable tools for observing these fast
Rabi oscillations by real-time X-ray images experimentally.
Fortunately, the coherent dynamical information for inner-shell
processes can be extracted from fluorescence spectra, based on
master equation approach.

As shown in the inset of Fig.~\ref{Ne_5}(a), we observe a triple-peak structure for the resonance fluorescence of the $1s^22s^22p^6\rightarrow 1s^12s^22p^63p^1$ transition, while only one central peak occurs for incoherent light pump. Considering the remarkable shift ($\approx 2$ eV) of the satellite line from the central peak for a laser intensity of $10^{18}\, {\rm W/cm^2}$, the triple-peak spectra provide a valuable tool for studying Rabi floppings and are expected to be detected via recording photons from spontaneous radiative decays~\cite{N. Rohringer1}. The next issue is related to the stability of the triple-peak structure against different experimental conditions, such as laser detuning and pulse shapes. We find that the triple-peak structure is stable. Specifically, we observe that the side peaks demonstrate asymmetric along with a red shift of the central peak for a red-detuned pump, as shown in Fig.~\ref{Ne_5}(b), whereas a Gaussian X-ray pulse with a FWHM duration of 15 fs yeids a broadened triple-peak structure compared to the flat-topped pulse, as shown in Fig.~\ref{Ne_5}(c). Note that we also observe a $\approx5$ eV broadening due to the extreme strong X-ray laser with an intensity of $10^{18}$ ${\rm W/cm^2}$, as shown in the guideline in the inset of Fig.~\ref{Ne_5}(b), which is normally referred to as power broadening as will be discussed in Sec.~\ref{real}. We expect that our discussions provide valuable insight for investigating inner-shell coherent dynamics in the upcoming experiments.

\subsection{Real coherent dynamics of a neon gas irradiated by an X-ray laser}\label{real}
\begin{figure}[h!]
\begin{tabular}{c}
\includegraphics[width=\linewidth]{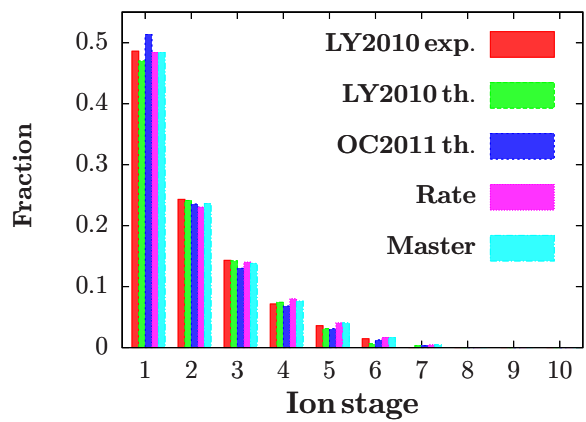}
\end{tabular}
\vspace{-2mm} \caption{Neon charge-state yields by a far
off-resonant beam with photon energy of 800 eV. Good agreements between different
theories~\cite{Young, O. Ciricosta} and experiments~\cite{Young}
indicate that coherence, induced by a far off-resonant laser beam,
plays a tiny role in the time evolution for the present experimental
conditions.}\label{compare}
\end{figure}
In former sections, coherence effects have been investigated through ultrafast dynamics of model systems coupled by X-ray pulses with different
temporal coherence. In this section, we will investigate coherence
effects beyond model systems, and take neon as examples for {\it real} dynamics of complex atoms
induced by an X-ray laser, based on master equation approach by
including thousands of atomic levels. The X-ray laser can sequentially
excite inner-shell electrons and creates a nonequilibrium state
decayed through a vast number of channels. To simplify our
simulations, we add the atomic orbitals gradually, and mainly focus
on dynamics of the lowest-lying orbitals of neon, such as $1s$,
$2s$, $2p$, $3n$ and $4n$, and inner-shell excited states $1s2n^2$, $1s2n3p$ and $1s2n4p$ as
well, which normally includes atomic levels up to an order of $10^3$
and spans the Hilbert subspace as basis states for {\it real}
dynamics of neon. Our selections are supported by the recent
R-matrix calculations for inner-shell electrons~\cite{Liu}. After
obtaining atomic data, we study the interplay between coherence-induced
effects and dissipations in X-ray-atom systems, based
on large scale simulations. We will verify to which extent coherence
demonstrates a dominant role in the ultrafast decayed nonequilibrium
system.

\paragraph{Off-resonant dynamics of neon in Young's experiment---}
\begin{figure}
\vspace{-13mm}
\begin{tabular}{c}
\includegraphics[width=1.25\linewidth]{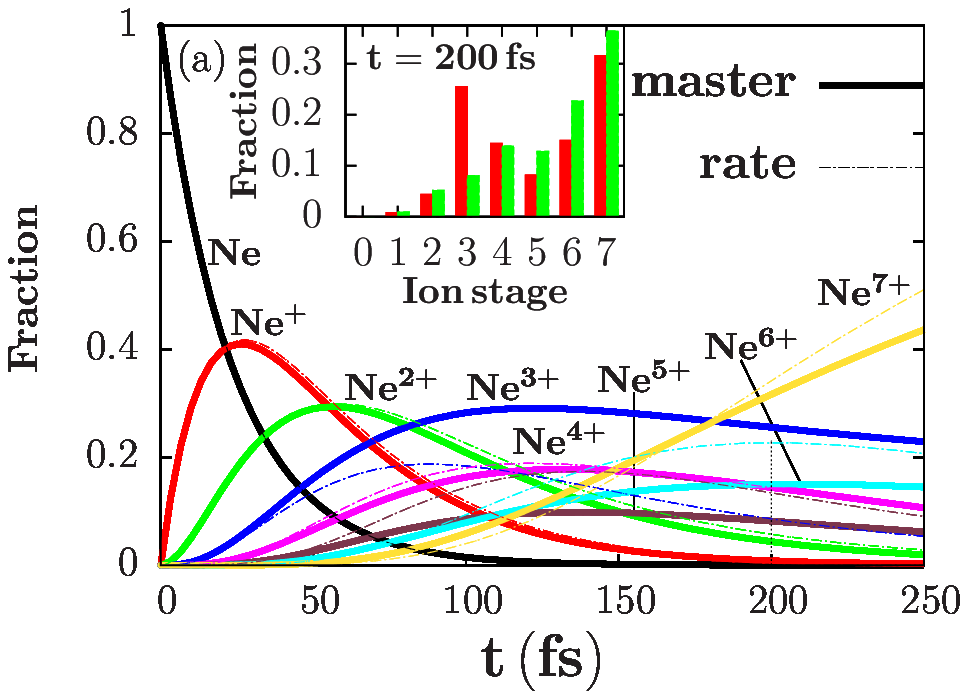}
\vspace{-18mm}
\\
\includegraphics[width=1.25\linewidth]{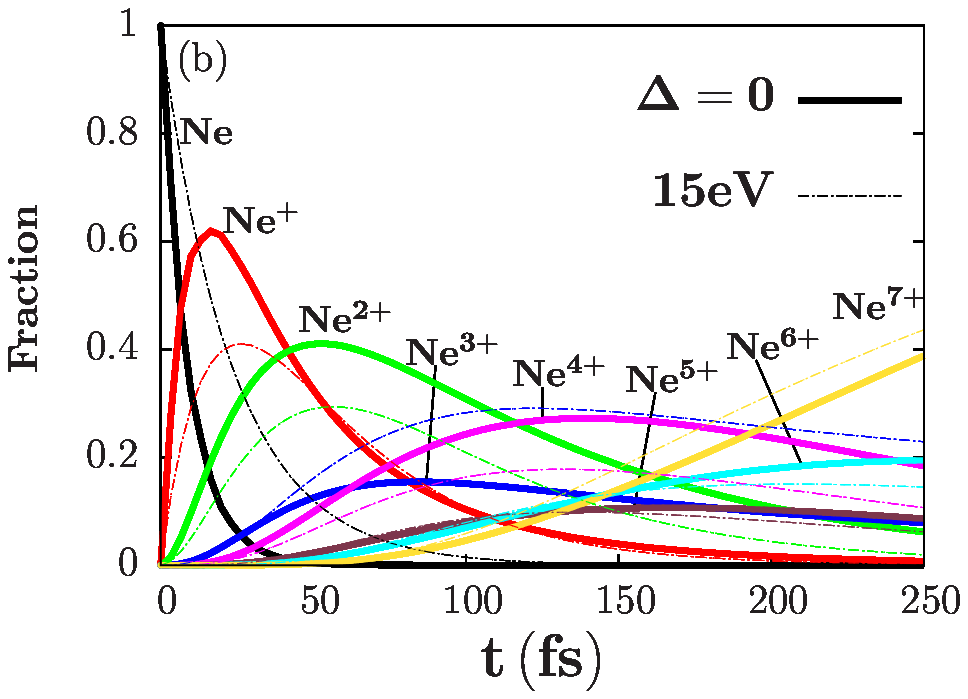}
\vspace{-18mm}
\\
\includegraphics[width=1.25\linewidth]{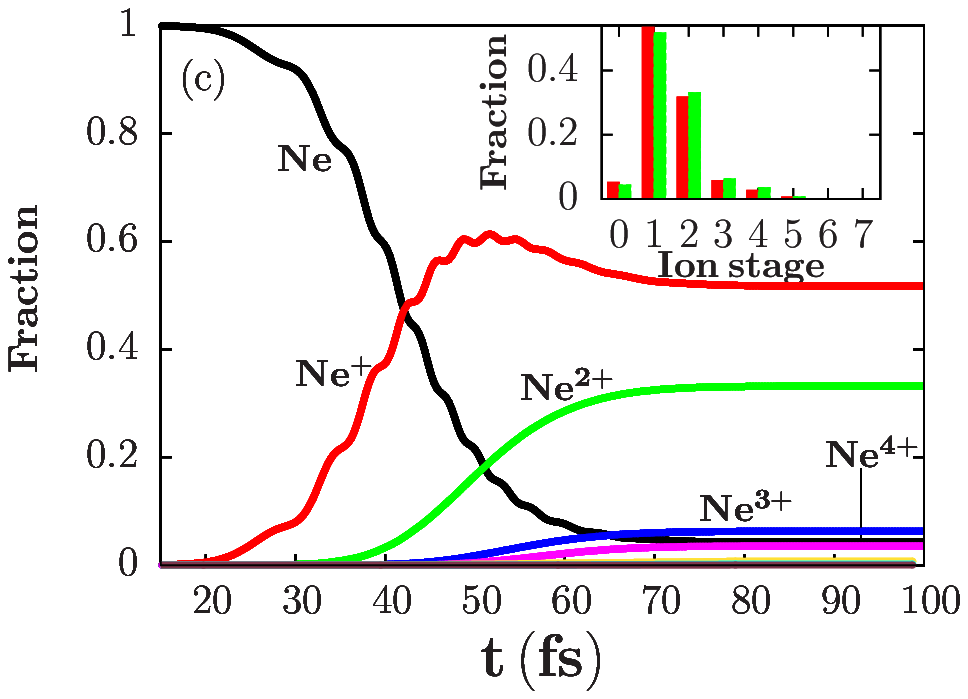}
\end{tabular}
\caption{Coherence-induced suppression on time evolution:
charge-state populations of neon induced by the X-ray laser with an
intensity of $2.5\times10^{17}\,{\rm W/cm^2}$ for different photon
energies: near-resonant excitation with a red detuning of 15 eV (a) and
resonant photoexcitation [(b),(c)] for $1s^22s^22p^6\rightarrow
1s^12s^22p^53p^1$ transition, and for different pulse shapes:
flat-topped [(a),(b)] and Gaussian pulses of a FWHM duration 25 fs
(c). Discrepancies between master equation and rate equation are the
results of coherence-induced Rabi oscillations and power-broadening
effects. {\bf Inset}: Fraction yields for different charge stages of
neon obtained by master equation (red) and rate equation (green) for
a pulse duration of 200 fs (a), and induced by a Gaussian pulse (red) of a
FWHM duration 25 fs and a flat-topped pulse (green) with the same
fluence (c).}\label{Ne_1}
\end{figure}

First we review the details of the X-ray free-electron-laser
experiment for neon in Ref.~\cite{Young}. In this experiment, X-ray pulses with photon energies of $800$, $1050$ and $2000$ eV
are injected into a neon gas in the atomic chamber, and they observe rapid
photoabsorptions of atomic gases in the fs timescale. As pointed out in the experiment, effects
due to particle collisions and photon scattering can be neglected,
since the neon gas is dilute and induced by pulses with photon
energies below 8 keV. Therefore, photoabsorptions are the dominant processes of neon
triggered by the X-ray beam in the ultra-intense,
short-wavelength regime. Here, we take photon energy of
$800$ eV as examples for investigating the influence of coherence on
the sequential multiphoton ionizations of the neon gas, where the
photon energy is far below resonant $1s \rightarrow 3p$ excitations
and it is referred to as off-resonant dynamics of neon. Remarkable agreements are obtained between different
theories~\cite{Young, O. Ciricosta} and experiments~\cite{Young}, as
shown in Fig.~\ref{compare}, which indicates that coherence plays a
tiny role in the time evolution of neon induced by a far
off-resonant X-ray pulse in the present experiments. For comparisons,
here we use both Gaussian and flat-topped pulses to simulate the
experimentally accessible dynamics of inner-shell electrons of neon,
and find that the charge-state distributions is insensitive to the
pulse shapes.

\paragraph{Near-resonant dynamics of inner-shell electrons of neon---}
In this paragraph, we study the time evolution of neon subjected to
a strong X-ray laser field with near-resonant photon energies, based
on large scale simulations by including energy levels in an order of
$10^3$. In Fig.~\ref{Ne_1}, we demonstrate charge state populations
as a function of time for a laser intensity $I_0=2.5\times10^{17}\,
{\rm W/cm^2}$, which is a typical free-electron-laser intensity in
the recent experiments, obtained via master equation [solid line in
Fig.~\ref{Ne_1}(a),(b),(c)] and rate equation approach [dashed line
in Fig.~\ref{Ne_1}(a)]. The photon energies are chosen near
resonant frequency relative to the $1s^22s^22p^6 \rightarrow
1s^12s^2p^63p^1$ transition, such as a red shift of 15 eV
[Fig.~\ref{Ne_1}(a),(b)] and the resonant case [Fig.~\ref{Ne_1}(b),(c)],
which are both in the photon-energy range of LCLS (800-2000 eV). We
find that coherence can suppress the multiphoton ionizations of neon
induced by the ultra-intense X-ray pulse. As shown in the inest of
Fig.~\ref{Ne_1}(a), for example, the charge state population of
${\rm Ne}^{3+}$ after a 200 fs evolution, is twice bigger than that
obtained by rate equation approach for incoherent pumping with an
intensity of $2.5\times10^{17}\,{\rm W/cm^2}$. The physical origin
of the discrepancy in the time evolution is the results of two
effects, which are neglected in the Einstein's rate equation. The
first one is the coherence in the inner-shell resonant absorption
processes, i.e $1s\rightarrow 3p$ excitations for $\rm Ne$ and
$1s\rightarrow2p$ for $\rm Ne^{3+}$, $\rm Ne^{4+}$ and $\rm
Ne^{5+}$. In contrast to monotonous changes in rate equation
approach, real physical processes between different energy levels
demonstrate a Rabi-flopping structure, due to the light-atom
coupling for the $1s \rightarrow 2p$ and $1s \rightarrow 3p$
transitions. The second one is the power broadening effects, due to
the extremely strong X-ray laser field, which can be up to an order
of a few eV at a laser intensity of $2.5\times10^{17}\, {\rm
W/cm^2}$, based on a two-level estimation $\sqrt{\gamma^2+
2\Omega^2}$ with $\gamma$ and $\Omega$ being spontaneous decay rate
and Rabi frequency, respectively. This point can also be verified
via fluorescence spectra, as shown in the inset of Fig.~\ref{Ne_5}(b).
Interestingly, we find that the red-shift case (relative to $1s
\rightarrow 3p$ transition) ionizes more electrons after 150 fs
evolution, since the $1s\rightarrow 2p$ excitations of inner-shell
electrons come into play for charge states ${\rm Ne}^{3+}$, ${\rm
Ne}^{4+}$ and ${\rm Ne}^{5+}$, whereas the resonant case ionizes
electrons fast in the early stage after subjected to the X-ray laser
field, since the resonant $1s \rightarrow 3p$ excitations dominate
photon absorptions of Ne in the first 50 fs, as shown in
Fig.~\ref{Ne_1}(b). The influence of temporal pulse shape on the
charge-state distributions is also discussed in Fig.~\ref{Ne_1}(c),
where comparison has been made between flat-topped (green) and
Gaussian pulses (red) in the inset. We remark here that the temporal
pulse shape has limited impact on the charge state distributions,
whose conclusion is consistent with those for incoherent
pulses~\cite{Young}.

\begin{figure}[h!]
\begin{tabular}{c}
\includegraphics[width=\linewidth]{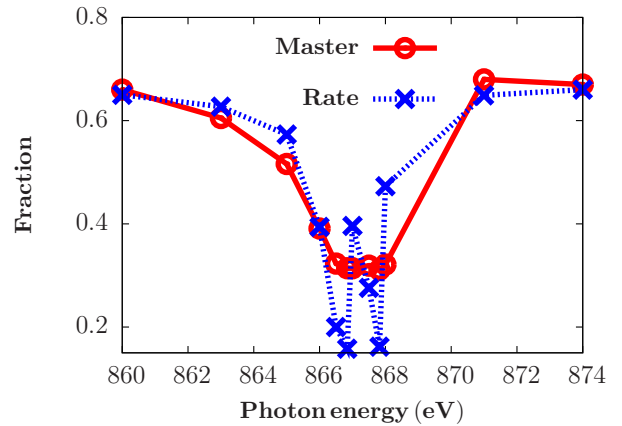}
\end{tabular}
\vspace{-2mm} \caption{Power broadening effects: Ne fraction in a neon gas subjected to a Gaussian X-ray pulse with an intensity of $2.5\times10^{17}\,{\rm W/cm^2}$ and a FWHM duration of 10 fs, for different photon energies, obtained by master equation (red solid) and rate equation approach (blue dashed). }\label{Ne_2}
\end{figure}

Next, we address the issue related to power broadening effects in
more details. Actually, the broadening effect, resulting in the line
shape of a dipole transition in an atom, is a basic feature for
describing electron motions and interactions with external fields.
The underlying mechanisms of broadening effects are diverse,
including natural broadening due to spontaneous decays of excited
states, Doppler broadening due to thermal motions of atoms,
collisional broadening due to collisions with other atoms or ions,
and Stark broadening due to energy shifts induced by an external
field. Here, we find that the dominant broadening effect for the
dilute neon gas in the LCLS experiment is the power broadening
($\approx 1$ eV) due to the extremely strong external laser, which
is up to 10 times bigger than those from Auger and spontaneous decay
processes ($\approx 0.1$ eV). Normally the rate equation approach
does not include the power broadening effects in the present
calculations, while in the master equation approach they are included
automatically. In Fig.~\ref{Ne_2}, we show the Ne fractions
subjected to a Gaussian X-ray laser with an intensity of
$2.5\times10^{17}\, {\rm W/cm^2}$ and a FWHM duration of 10 fs,
where two local minima, obtained from rate equation (blue dashed),
correspond to resonant excitations for $1s^22s^22p^6\rightarrow
1s^12s^22p^63p^1$ and $1s^22s^22p^6\rightarrow 1s^12s^22p^64p^1$
transitions, respectively. However, the oscillating structures for
resonant excitations at different photon energies are smoothen in
the results obtained from master equation approach (red circle),
due to power broadening effects. Note here that the oscillating
structures can occur at other photon energies associated with
$1s\rightarrow np\, (n\geq5)$ transitions, but are beyond the
spectrum resolution in the present experiments and an improved
sharp-band X-ray laser is needed in the future.

\section{Summary and Outlook}
In conclusion, we establish a general method for describing coherent
dynamics of X-ray-matter systems in the framework of master equation
approach. We take dilute atomic gases as examples for discussing
{\it real} coherent dynamics of the rapidly decayed X-ray-matter system,
based on a thousand state atomic master equation approach, by
including coherent pump and incoherent relaxations due to
spontaneous and Auger decays. We find that coherence can suppress
the sequential single-photon ionizations of a neon gas in the
ultra-intense X-ray field, compared to the rate equation approach,
and the physical reason can attribute to the coherence-induced Rabi
oscillations and power broadening effects, which are both neglected
in the Einstein's rate equation. We also find that single-photon
ionizations for both outer- and inner-shell electrons dominate the
absorptions of a neon gas for the recently typical experiments with
a laser intensity of $\approx10^{18}$ ${\rm W/cm^2}$, irrespective
of coherence. A typical feature of coherent evolution
of inner-shell electrons is Rabi oscillations with a frequency in
the order of $10^{15}$ Hz, which is beyond the current experimental
resolutions in the time domain. Instead, we discuss resonance
fluorescence spectra for possible experimental implementations for
coherent dynamics of inner-shell electrons.


With the quick development of free-electron lasers, temporal
coherence of the X-ray laser is improved experimentally, which
provides the possibility for studying time-resolved coherent
phenomena in atoms, molecules and solid-state materials~\cite{M.
Drescher}. There are still a number of open issues referred to
coherent dynamics of complex systems irradiated by an intense X-ray
laser. For the methods described here, we are able to discuss
hollow-atom signatures and its corresponding coherent dynamics, where
the competition between the double- and single-hole generations in
coherent evolution is still unclear. Moreover, we can also
investigate multiphoton processes of atoms and molecules in gases
and solid-dense matters, such as two-photon KK-shell transitions,
where the issues are related to non-sequential double ionizations
and electron-electron correlations. Finally, in a dense hot plasma
environment generated by dense gases or solid-state materials
induced by an X-ray laser, the environment is much more complicated,
both radiative and particle-collision processes should be taken into
account for understanding the phenomena of the system.

\begin{acknowledgments}
We acknowledge useful discussions with Z.-X. Zhao and H.-Y. Sun.
This work is supported by the National Basic Research Program of
China (973 Program) under Grant No. 2013CB922203, the National
Natural Science Foundation of China under Grant Nos. 11274383,
11304386, 11104350 and 11204376.
\end{acknowledgments}


\begin{references}
\bibitem{RPL1} L. Di\'osi and L. Ferialdi, Phys. Rev. Lett. {\bf 113}, 200403 (2014).
\bibitem{RPL2} F. Kadi, T. Winzer, E. Malic, A. Knorr, F. G\"otfert, M. Mittendorff, S. Winnerl, and M. Helm, Phys. Rev. Lett. {\bf 113}, 035502 (2014).
\bibitem{PRL3} T. Ambj\"ornsson, S. K. Banik, O. Krichevsky and R. Metzler, Phys. Rev. Lett. {\bf 97}, 128105 (2006).
\bibitem{PRL4} Y.-J. Wei, Y. He, Y.-M. He, C.-Y. Lu, J.-W. Pan, C. Schneider, M. Kamp, S. H\"ofling, D. P. S. McCutcheon and A. Nazir, Phys. Rev. Lett. {\bf 113}, 097401, (2014).
\bibitem{J. Cirac} A. C. Pflanzer, O. Romero-Isart and J. I. Cirac, Phys. Rev. A {\bf 86}, 013802 (2012).
\bibitem{A. Imamoglu} A. Imamoglu, J. E. Field and S. E. Harris, Phys. Rev. Lett. {\bf 64}, 1154 (1991).
\bibitem{S. M. Cavaletto} S. M. Cavaletto, C. Buth, Z. Harman, E. P. Kanter, S. H. Sourthworth, L. Young and C. H. Keitel, Phys. Rev. A {\bf 86}, 033402 (2012).
\bibitem{Seres} J. Seres, E. Seres A. J.Verhoef, G. Tempea, C. Streli, P.Wobrauschek, V. Yakovlev, A. Scrinzi, C. Spielmann and F. Krausz, Nat. (London) {\bf 433}, 596 (2005).
\bibitem{Maiman} T. H. Maiman, Nat. (London) {\bf 187}, 493 (1960).
\bibitem{Fano} U. Fano, Phys. Rev. {\bf 124}, 1866 (1961).
\bibitem{Annett} J. F. Annett, Superconductivity, Superfluid and Condensates, (Oxford University Press 2004).
\bibitem{BEC} E. A. Cornell and C. E. Wieman, Rev. Mod. Phys. {\bf 74}, 875 (2002); W. Ketterle, Rev. Mod. Phys. {\bf 74}, 1131 (2002).

\bibitem{Harris} S. E. Harris, Phys. Today {\bf 50}, 36 (1997); A. Imamoglu and S. E. Harris, Opt. Lett. {\bf 14}, 1344 (1989).
\bibitem{Plaja} L. Plaja, R. Torres, A. Za\"ir, Attosecond Physics: Attosecond Measurements and Control of Physical Systems, (Springer London Press 2013).
\bibitem{Harris2} K.-J. Boller, A. Imamoglu, and S.E. Harris., Phys. Rev. Lett. {\bf 66}, 2593 (1991).
\bibitem{SF-MI} M. Greiner, O. Mandel, T. Esslinger, T. W. H\"ansch and I. Bloch, Nat. (London) {\bf 415}, 39 (2002).
\bibitem{P. Emma_2010}P. Emma {\it et al.}, Nat. Photon. {\bf 4}, 641 (2010).
\bibitem{F. Lehmkuhler} F. Lehmk\"uhler {\it et al.} Sci. Rep. {\bf 4}, 5234 (2014).
\bibitem{B. Nagler} B. Nagler {\it et al.}, Nat. Phys. {\bf 5}, 693 (2009).
\bibitem{Young} L. Young, E. P. Kanter, B. Kr\"assig, Y. Li, A. M. March, S. T. Pratt, R. Santra, S. H. Southworth, N. Rohringer, L. F. DiMauro, G. Doumy, C. A. Roedig, N. Berrah, L. Fang, M. Hoener, P. H. Bucksbaum, J. P. Cryan, S. Ghimire, J. M. Glownia, D. A. Reis, J. D. Bozek, C. Bostedt and M. Messerschmidt, Nat. (London) {\bf 466}, 56 (2010).
\bibitem{M. Hoener} M. Hoener, L. Fang, O. Kornilov, O. Gessner, S. T. Pratt, M. G\"uhr, E. P. Kanter, C. Blaga, C. Bostedt, J. D. Bozek, P. H. Bucksbaum, C. Buth, M. Chen, R. Coffee, J. Cryan, L. DiMauro, M. Glownia, E. Hosler, E. Kukk, S. R. Leone, B. McFarland, M. Messerschmidt, B. Murphy, V. Petrovic, D. Rolles and N. Berrah, Phys. Rev. Lett. {\bf 104}, 253002 (2010).
\bibitem{J. Cryan} J. P. Cryan, J. M. Glownia, J. Andreasson, A. Belkacem, N. Berrah, C. I. Blaga, C. Bostedt, J. Bozek, C. Buth, L. F. DiMauro, L. Fang, O. Gessner, M. Guehr, J. Hajdu, M. P. Hertlein, M. Hoener, O. Kornilov, J. P. Marangos, A. M. March, B. K. McFarland, H. Merdji, V. S. Petrovi\'c, C. Raman, D. Ray, D. Reis, F. Tarantelli, M. Trigo, J. L. White, W. White, L. Young, P. H. Bucksbaum, and R. N. Coffee, Phys. Rev. Lett. {\bf 105}, 083005 (2010).
\bibitem{G. Doumy} G. Doumy, C. Roedig, S.-K. Son, C. I. Blaga, A. D. DiChiara, R. Santra, N. Berrah, C. Bostedt, J. D. Bozek, P. H. Bucksbaum, J. P. Cryan, L. Fang, S. Ghimire, J. M. Glownia, M. Hoener, E. P. Kanter, B. Kr\"assig, M. Kuebel, M. Messerschmidt, G. G. Paulus, D. A. Reis, N. Rohringer, L. Young, P. Agostini and L. F. DiMauro, Phys. Rev. Lett. {\bf 106}, 083002 (2011).
\bibitem{H. Thomas} H. Thomas, A. Helal, K. Hoffmann, N. Kandadai, J. Keto, J. Andreasson, B. Iwan, M. Seibert, N. Timneanu, J. Hajdu, M. Adolph, T. Gorkhover, D. Rupp, S. Schorb, T. M\"oller, G. Doumy, L. F. DiMauro, M. Hoener, B. Murphy, N. Berrah, M. Messerschmidt, J. Bozek, C. Bostedt and T. Ditmire1, Phys. Rev. Lett. {\bf 108}, 133401 (2012).
\bibitem{N. Rohringer1} N. Rohringer {\it et al.}, Nat. (London) {\bf 481}, 488 (2012).
\bibitem{B. Rudek} B. Rudek {\it et al.}, Nat. Photon. {\bf 6}, 858 (2012).
\bibitem{S. M. Vinko} S. M. Vinko {\it et al.}, Nat. (London) {\bf 482}, 59 (2012).
\bibitem{H. Fukuzawa} H. Fukuzawa {\it et al.}, Phys. Rev. Lett. {\bf 110}, 173005 (2013); H. Fukuzawa {\it et al.}, Phys. Rev. Lett. {\bf 111}, 043001 (2013).
\bibitem{N. Rohringer2} N. Rohringer and R. Santra, Phys. Rev. A {\bf 76}, 033416 (2007).
\bibitem{O. Ciricosta} O. Ciricosta,, H.-K. Chung, R. W. Lee and J. S. Wark, Hi. Ene. Dens. Phys. {\bf 7}, 111 (2011).
\bibitem{W.-J. Xiang} W.-J. Xiang, C. Gao, Y.-S. Fu, J.-L. Zeng and J.-M. Yuan, Phys. Rev. A {\bf 86}, 061401(R) (2012).
\bibitem{Robert} R. H\"oppner, Eugenio Rold\'an, and G. J. de Valc\'arcel, Am. J. Phys. {\bf 80}, 882 (2012).
\bibitem{SASE} R. Bonifacio, C. Pellegrini and L. M. Narducci, Opt. Commun. {\bf 50}, 373 (1984).
\bibitem{P. Emma} P. Emma, K. Bane, M. Cornacchia, Z. Huang, H. Schlarb, G. Stupakov, and D. Walz, Phys. Rev. Lett. {\bf 92}, 074801 (2004).
\bibitem{Y. Ding} Y. Ding, A. Brachmann, F.-J. Decker, D. Dowell, P. Emma, J. Frisch, S. Gilevich, G. Hays, Ph. Hering, Z. Huang, R. Iverson, H. Loos, A. Miahnahri, H.-D. Nuhn, D. Ratner, J. Turner, J. Welch, W. White, and J. Wu, Phys. Rev. Lett. {\bf 102}, 254801 (2009).
\bibitem{N. R. Thompson} N. R. Thompson and B. W. J. McNeil, Phys. Rev. Lett. {\bf 100}, 203901 (2008).
\bibitem{Y. Wang} Y. Wang {\it et al.}, Nat. Photon. {\bf 2}, 94 (2008).
\bibitem{J. Zhao} J. Zhao {\it et al.}, Opt. Express {\bf 16}, 3546 (2008).
\bibitem{K. Lan} K. Lan, E. Fill, and J. Meyer-Ter-Vehn, Laser Part. Beams {\bf 22}, 261 (2004).
\bibitem{S. Jacquemot} S. Jacquemot, K. T. Phuoc, A. Rousse, and S. Sebban, X-ray Lasers 2006 (Springer, New York, 2007).
\bibitem{N. Rohringer3} N. Rohringer and R. London, Phys. Rev. A {\bf 80}, 013809 (2009).
\bibitem{S. Ackermann} S. Ackermann {\it et al.}, Phys. Rev. Lett. {\bf 111}, 114801 (2013).
\bibitem{Kenneth C. Kulander} K. C. Kulander, Phys. Rev. A {\bf 35}, 445 (1987).
\bibitem{M. Lewenstein} M. Lewenstein, P. Balcou, M. Y. Ivanov, A. L'Huillier and P. B. Corkum, Phys. Rev. A {\bf 49}, 2117 (1993).
\bibitem{M. Protopapas} M. Protopapas, C. H. Keitel and P. L. Knight, Rep. Prog. Phys. {\bf 60}, 389 (1997).
\bibitem{Z.-X. Zhao} S. Patchkovskii1, Z.-X. Zhao and T. Brabec and D. M. Villeneuve, Phys. Rev. Lett. {\bf 97}, 123003 (2006).
\bibitem{J. Zhao_2008} J. Zhao and Z.-X Zhao, Phys. Rev. A {\bf 78}, 053414 (2008).
\bibitem{B. Zhang} B. Zhang, J.-M. Yuan and Z.-X. Zhao, Phys. Rev. Lett. {\bf 111}, 163001 (2013).
\bibitem{H.-P. Breuer} H.-P. Breuer and F. Petruccione, The theory of open quantum systems, (Oxford University Press 2002).
\bibitem{B. R. Mollow} B. R. Mollow, Phys. Rev. A {\bf 12}, 1919 (1975).
\bibitem{R. Brewer} R. Brewer and E. L. Hahn, Phys. Rev. A {\bf 11}, 1641 (1975).
\bibitem{P. Zoller} R. Dum, P. Zoller and H. Ritsch, Phys. Rev. A {\bf 45}, 4879 (1992).
\bibitem{P. Zoller_1993} P. Marte, R. Dum, R. Taieb and P. Zoller, Phys. Rev. A {\bf 47}, 1378 (1993).
\bibitem{X.-M. Hu} X.-M. Hu and J.-P. Zhang, J. Phys. B {\bf 37}, 345 (2004).
\bibitem{M. Lax} M. Lax, Phys. Rev. {\bf 129}, 2342 (1963).
\bibitem{J. H. Eberly} J. H. Eberly, C. V. Kunasz and K. Wodkiewicz, J. Phys. B {\bf 13}, 217 (1980).
\bibitem{M. Florjanczyk} M. Florjanczyk, K. Rzazewski and J. Zakrzewski, Phys. Rev. A {\bf 31}, 1558 (1985)
\bibitem{M. Wilkens} M. Wilkens and K. Rzazewski, Phys. Rev. A {\bf 40}, 3164 (1989).
\bibitem{R. J. Glauber} R. J. Glauber, In quantum optics and electronics, (Gordon, and Breach, New York, 1965).
\bibitem{Li_2008}  Y.-Q. Li, J.-H. Wu, Y. Hou and J.-M. Yuan, J. Phys. B {\bf 4},145002 (2008).
\bibitem{Gao_2013} C. Gao, J.-L. Zeng, Y.-Q. Li, F.-T. Jin, J.-M. Yuan, Hi. Ene. Dens. Phys. {\bf 9}, 583 (2013).
\bibitem{Gao_2} C. Gao, F.-T. Jin, J.-L. Zeng and J.-M. Yuan, New J. Phys. {\bf 15}, 015022 (2013).
\bibitem{J. Zeng} J.-L. Zeng, P.-F. Liu, W.-J. Xiang, J.-M. Yuan, Phys. Rev. A {\bf 87}, 033419 (2013).
\bibitem{S. Palaniyappan} S. Palaniyappan, {\it et al.}, Phys. Rev. Lett. {\bf 94}, 243003 (2005).
\bibitem{W. Ackermann} W. Ackermann, {\it et al.}, Nat. Photon. {\bf 1}, 336 (2007).
\bibitem{A. A. Sorokin} A. A. Sorokin, {\it et al.}, Phys. Rev. Lett. {\bf 99}, 213002 (2007).
\bibitem{Liu} Y.-P. Liu, J.-L. Zeng and J.-M. Yuan, J. Phys. B: At. Mol. Opt. Phys. {\bf 46}, 145002 (2013).
\bibitem{M. Drescher} M. Drescher, M. Hentschel, R. Kienberger, M. Uiberacker, V. Yakovlev, A. Scrinzi, Th. Westerwalbesloh, U. Kleineberg,
U. Heinzmann and F. Krausz, Nat. (London) {\bf 419}, 803 (2002).
 \end{references}
\end{document}